\documentclass[11pt]{article}
\usepackage[utf8x]{inputenc}
\usepackage{amssymb}
\usepackage{amsmath}
\usepackage{amscd}
\usepackage{latexsym}
\usepackage{graphics}
\usepackage{color}
\usepackage{mathtools}
\usepackage{cite}
\usepackage{booktabs}
\usepackage{tikz}
\usepackage{hyperref}
\usepackage{enumerate}
\usetikzlibrary{arrows}
\input xy
\xyoption{all}

\catcode`@=11

\numberwithin{equation}{section}
\catcode`@=12

\def\a{\alpha}
\def\b{\beta}
\def\g{\gamma}

\def\e{\epsilon}
\def\l{\lambda}

\def\m{\mu}
\def\n{\nu}

\renewcommand{\e}{\,\mathrm{e}\,}

\newcommand{\im}{\,\mathrm{i}\,}
\newcommand{\diff}{\mathrm{d}}

\newcommand{\R}{{\mathbb{R}}}
\newcommand{\N}{{\mathbb{N}}}
\newcommand{\C}{{\mathbb{C}}}

\newcommand{\HH}{{\mathbb{H}}}

\newcommand{\ca}{{\cal{A}}}
\newcommand{\cf}{{\cal{F}}}

\newcommand{\tr}{{\rm tr}}

\newcommand{\End}{\mathrm{End}}

\def\>{\rangle}
\def\<{\langle}
\def\+{\dagger}
\def\={\ =\ }

\newcommand{\lb}{\left(}
\newcommand{\rb}{\right)}
\newcommand{\com}[2]{\left[#1, #2\right]}

\newcommand{\urm}{\mathrm{U}}
\newcommand{\urmL}{\mathfrak{u}}
\newcommand{\surm}{\mathrm{SU}}
\newcommand{\surmL}{\mathfrak{su}}
\newcommand{\sorm}{\mathrm{SO}}
\newcommand{\sprm}{\mathrm{Sp}}
\newcommand{\sormL}{\mathfrak{so}}
\newcommand{\sprmL}{\mathfrak{sp}}
\newcommand{\Ncal}{\mathcal{N}}
\newcommand{\Ocal}{\mathcal{O}}
\newcommand{\Pcal}{\mathcal{P}}
\newcommand{\Qcal}{\mathcal{Q}}
\newcommand{\Scal}{\mathcal{S}}
\newcommand{\Tcal}{\mathcal{T}}
\newcommand{\Xcal}{\mathcal{X}}
\newcommand{\Ycal}{\mathcal{Y}}
\newcommand{\Zcal}{\mathcal{Z}}
\newcommand{\glrm}{\mathrm{GL}}
\newcommand{\glrmL}{\mathfrak{gl}}
\newcommand{\slrm}{\mathrm{SL}}
\newcommand{\slrmL}{\mathfrak{sl}}

\newcommand{\Ad}{\mathrm{Ad}}
\newcommand{\ad}{\mathrm{ad}}
\newcommand{\gfrak}{\mathfrak{g}}
\newcommand{\hfrak}{\mathfrak{h}}
\newcommand{\mfrak}{\mathfrak{m}}
\newcommand{\Aa}{\mathbb{A}}

\newcommand{\Ggauge}{\widehat{\mathcal{G}}}
\newcommand{\ggauge}{\widehat{\mathfrak{g}}}
\newcommand{\Int}{\mathrm{Int}}
\newcommand{\bg}{\boldsymbol{g}}
\newcommand{\bw}{\boldsymbol{\omega}}
\newcommand{\bJ}{\boldsymbol{J}}
\newcommand{\cm}{\mathcal{M}}
\newcommand{\mfd}[1]{M^{#1}}

\newcommand{\Mcal}{\mathcal{M}}
\newcommand{\rank}{\mathrm{rk}}
\newcommand{\diag}{\mathrm{diag}}
\begin{document}
\begin{titlepage}
\setcounter{page}{0}
\begin{flushright}
UWTHPH-2017-28 
\end{flushright}

\vskip 2cm

\begin{center}

{\Large\bf Instantons on Calabi-Yau and hyper-Kähler cones
}

\vspace{15mm}

{\large Jakob C. Geipel${}^{1}$} , \ {\large Marcus Sperling${}^{2}$} 
\\[5mm]
\noindent ${}^1$ \emph{Institut f\"ur Theoretische Physik, Leibniz 
Universit\"at Hannover}\\
\emph{Appelstra\ss e 2, 30167 Hannover, Germany}\\
{Email: {\tt jakob.geipel@itp.uni-hannover.de}}
\\[5mm]
\noindent ${}^{2}$ \emph{Fakultät für Physik, Universität Wien}\\
\emph{Boltzmanngasse 5, 1090 Wien, Austria}\\
Email: {\tt marcus.sperling@univie.ac.at}
\\[5mm]

\vspace{15mm}

\begin{abstract}
The instanton equations on vector bundles over Calabi-Yau and hyper-Kähler 
cones can be reduced to matrix equations resembling Nahm's equations. 
We complement the discussion of Hermitian Yang-Mills (HYM) equations on 
Calabi-Yau cones, based on regular semi-simple elements, by a 
new set of (singular) boundary conditions which have a known instanton solution 
in one direction. 
This approach extends the classic results of Kronheimer by probing a relation 
between generalised Nahm's equations and nilpotent pairs/tuples. 
Moreover, we consider quaternionic instantons on hyper-Kähler cones over generic 
3-Sasakian manifolds and study the HYM moduli spaces arising in this set-up, 
using the fact that their analysis can be traced back to the 
intersection of three Hermitian Yang-Mills conditions.
\end{abstract}

\end{center}

\end{titlepage}

{\baselineskip=12pt
\tableofcontents
}

\section{Introduction}
Instantons and (hyper-)Kähler geometry are both interesting subjects for 
physicists as well as mathematicians, and important results in mathematical physics 
have been derived by studying the structure of moduli spaces of certain gauge connections.

The geometry of hyper-Kähler manifolds is in itself very restrictive and there is to this day 
no explicit compact hyper-Kähler metric known. Nonetheless, the classification 
of compact hyper-Kähler spaces is understood \cite{Gross:2003} and yields four 
classes: two series of $\mathrm{K}3^n$ and generalised Kummer varieties, as 
well as two exceptional examples by O'Grady.
In contrast, many examples of (non-compact) hyper-Kähler spaces arise as moduli 
spaces of gauge theory problems: moduli spaces of instantons, monopoles, or 
Hitchin equations, to name a few.
As it turns out \cite{Hitchin:1986ea}, 
hyper-Kähler geometry is intimately related to supersymmetry. Various moduli 
spaces of supersymmetric vacua are hyper-Kähler: including 
manifestations of the so-called hyper-Kähler quotient and of 
hyper-Kähler cones \cite{Gibbons:1998xa,deWit:1998zg}. 

Instantons on $4$-manifolds, meaning 
(anti-)self-dual connections, led to an improved understanding via the concept 
of Donaldson invariants \cite{Donaldson:1983}. On the other hand, 
$4$-dimensional Euclidean instantons are vital for non-perturbative effects in 
quantum field theory and string theory.
The generalisation of the notion of instantons to higher dimensions has been 
first proposed by \cite{Corrigan:1982th}. In particular, instantons can be 
defined on any manifold with a $G$-structure.
Suppose $G=\urm(m)$, such that the (compact) $2m$-dimensional manifold is 
endowed with a Kähler structure. In this case, the instanton equations are known 
as Hermitian Yang-Mills equations and have a deep geometric interpretation in 
form of the Donaldson-Uhlenbeck-Yau theorem 
\cite{Donaldson:1985,Uhlenbeck:1986}.
In contrast, for holonomy $G=\sprm(m)$, the $4m$-dimensional manifold is hyper-Kähler 
and the notion of instantons on such spaces has been proposed 
by \cite{Salamon:1988,Nitta:1988}. The generalised Ward correspondence 
\cite{Bartocci:2004}, which 
relates \emph{quaternionic instantons} on a hyper-Kähler space $\mfd{4m}$ with 
some 
holomorphic vector bundle on the twistor space of $\mfd{4m}$, again provides a 
relation between gauge theory and holomorphic bundles.  

Considering compactifications of 10-dimensional heterotic string theory which 
preserve 
$\Ncal{=}1$ supersymmetry in 4 dimensions, one has to satisfy the so-called 
BPS equations, 
which contain an instanton equation on the internal $G$-structure manifold. 
Unfortunately, for compact Calabi-Yau or compact hyper-Kähler spaces, 
explicit metrics are not known, but one can resort to cone constructions as a 
testing ground. The underlying base for a hyper-Kähler cone is a 3-Sasakian 
space, while a Calabi-Yau cone starts from a Sasaki-Einstein base. 
Instantons on certain conical extensions of $G$-manifolds have been 
considered, for instance, in 
\cite{Harland:2009yu,Harland:2010ix,Bauer:2010fia,Haupt:2011mg,Gemmer:2011cp,
harland_noelle,ivanova_popov,Bunk:2014kva,Bunk:2014coa,Sperling:2015sra,
Haupt:2015wdq}. In all the references, the instanton equations have been 
reduced to a set of matrix equations by a certain equivariant ansatz. 
The resulting matrix equations for Calabi-Yau cones over arbitrary 
Sasaki-Einstein manifolds have been discussed in \cite{Sperling:2015sra} for 
one choice of boundary conditions. The 
aim of the present paper is twofold: firstly, to extend the discussion on the 
Calabi-Yau cones by considering different boundary conditions, which appear to 
be more physically relevant. Secondly, to generalise and extend this analysis 
to the matrix equations resulting from the \emph{quaternionic instanton 
equation} on hyper-Kähler cones over arbitrary 3-Sasakian manifolds.

Interestingly, the instanton matrix equations resulting from the equivariant 
reduction can be viewed as \emph{generalised Nahm's equations}, called 
\emph{Nahm-type equations} in \cite{Sperling:2015sra}. Recalling the prominent 
role of Nahm's equations and nilpotent orbits for BPS boundary conditions for 
4-dimensional $\Ncal{=}4$ super Yang-Mills 
theories \cite{Gaiotto:2008sa,Gaiotto:2008ak}, the constructions of 
4-dimensional $\Ncal{=}1$ theories by compactifying 6-dimensional theories 
\cite{Xie:2013gma,Heckman:2016xdl} or assigning $1/4$ BPS boundary 
conditions\cite{Hashimoto:2014vpa,Hashimoto:2014nwa} on 4-dimensional 
$\Ncal{=}4$ super-Yang-Mills led to the appearance of generalised Nahm's 
equations. These are in fact dimensional reductions of 6-dimensional Hermitian 
Yang-Mills equations and the moduli space of the generalised Nahm's 
equations will be related to orbits of commuting nilpotent pairs.

The outline of the article is as follows: in Section \ref{sec:geometry} we 
briefly recall the geometry of Sasaki-Einstein and 3-Sasakian spaces as well as 
their metric cones. Section 
\ref{sec:comments_moduli_space} is devoted to a description of the moduli space 
of quaternionic instantons, starting with the space of connections, then 
showing a reformulation of the $\sprm(m)$-instanton moduli space as 
intersection of various $\surm(2m)$-moduli spaces. At the end of this 
section, we specialise to an 
equivariant ansatz which reduces the instanton equation to a set of matrix 
equations. 
In Section \ref{sec:instanton_cy-cone} we treat the instanton matrix equations 
on the Calabi-Yau cone and show that, depending on the boundary conditions, the 
moduli space relates to different ``diagonal''  complex coadjoint orbits. In 
particular, the choice of singular boundary conditions for the generalised 
Nahm's equations will naturally lead to orbits of tuples of commuting nilpotent 
elements.
In Section \ref{sec:instanton_hk-cone}, we subsequently extend this study to 
the instantons on hyper-Kähler cones. Finally, Section \ref{sec:conclusion} 
concludes.
Appendix \ref{app:details} provides some technical details.

\section{Geometry}
\label{sec:geometry}
In this section we review the definitions and relevant properties of 
Sasaki-Einstein and 3-Sasakian manifolds as well as the geometric structure of 
their metric cones. For details we refer to 
\cite{boyer_galicki,boyer_galicki_mann,hitchin_hyperkaehler,BFGK,harland_noelle} and the references therein.

\paragraph{Sasaki-Einstein manifolds.}

Sasaki manifolds are the odd-dimensional analogues of K\"ahler manifolds in the sense that a Riemannian manifold $(M^{2n+1},g)$ is \emph{Sasakian}
if and only if its metric cone is a Kähler manifold. That is, the metric cone is 
a complex manifold with closed Kähler form $\Omega(X,Y) \coloneqq 
g(X,JY)$, or, equivalently, a manifold whose holonomy group is  contained in the unitary group $U(n+1)$.

An equivalent definition (see \cite{boyer_galicki}) is that of a manifold which
admits a Killing vector field $\xi$ of unit length such that the type-(1,1) vector 
field $\Phi(X) \coloneqq \nabla^{\mathrm{LC}}_X \xi$ satisfies
\begin{equation}
 (\nabla_{X}^{\mathrm{LC}} \Phi) (Y) = g(\xi, Y) X - g(X,Y) \xi
 \end{equation}
for all vector fields $X$ and $Y$ on $M$. The vector field $\xi$ is referred
to as \emph{characteristic} or \emph{Reeb vector field}, and Sasakian manifolds
are a subclass of metric contact structures. 

Denoting by $\eta$ the 1-form dual to $\xi$, one can consider the 
\emph{Reeb foliation} along the characteristic vector field, given by 
the subbundle $\mathcal{D}\coloneqq \mathrm{ker}(\eta)$. This yields 
transverse Kähler spaces of real dimensions $2n$, and the corresponding 
Kähler form $\omega$
follows from the relation $\diff \eta =2 \omega$.

A \emph{Sasaki-Einstein manifold} $(\mfd{2n+1},g,\xi)$ is a Sasakian manifold 
whose metric is 
additionally Einstein, which implies that the metric cone is \emph{Calabi-Yau},
i.e.\ a \emph{Ricci-flat}
Kähler manifold. The latter is equivalent to a manifold with 
special holonomy contained in $\surm(n{+}1) \subset \urm(n{+}1) \subset 
\sorm(2n{+}2)$.

\paragraph{3-Sasakian manifolds.}
A \emph{3-Sasakian manifold} is a Riemannian manifold $(\mfd{4m+3},g)$ of 
real dimension $4m+3$ which admits a triplet of Sasaki structures such that 
their
characteristic vector fields $\xi_{\a}$ are orthogonal, $g(\xi_{\a}, 
\xi_{\b})=\delta_{\a\b}$, and satisfy the $\surm(2)$ commutation relations,
\begin{equation}
\com{\xi_{\a}}{\xi_{\b}}=2\epsilon_{\a\b}^{\phantom{\a\b} \g}\xi_{\g}.
\end{equation}
Note that the existence of this  triple of characteristic vector 
fields implies  a whole $\C P^1$ family of those structures.
Moreover, it can be shown that every 3-Sasakian manifold $\mfd{4m+3}$ is 
automatically Einstein and that its structure group is $\sprm(m)$ (see e.g. 
\cite{boyer_galicki} and the references therein). An alternative definition for 
3-Sasakian manifolds is that of being a manifold $\mfd{4m+3}$ such that the 
metric 
cone is hyper-Kähler, i.e.\ 
its Riemannian holonomy is contained in $\sprm(m{+}1)$. Consequently, a 
prototype 
of 3-Sasakian manifolds are the homogeneous spaces
$\sprm(m{+}1)/\sprm(m) \cong S^{4m+3}$, the \emph{squashed spheres}. 
The squashed spheres are $\surm(2)$-bundles, while all other homogeneous 
3-Sasakian manifolds are $\sorm(3)$-bundles over quaternionic spaces (see e.g. 
\cite{boyer_galicki}). Another well-known example is the seven-dimensional 
Aloff-Wallach space $X_{1,1} \cong \surm(3)/\urm(1)_{1,1}$.

\paragraph{Hyper-Kähler cones.}
By definition, a manifold is 3-Sasakian if its metric cone is 
\emph{hyper-Kähler}, 
i.e.\ it admits a triplet of covariantly constant complex structures $J_1$, 
$J_2$, and $J_3$ satisfying the quaternionic relations
\begin{align}
\label{eq:quat_relations}
J_\alpha J_\beta = -\delta_{\alpha \beta}\, \mathrm{id} 
+ \epsilon_{\alpha\beta}^{\phantom{\a\b} \gamma} J_\gamma \; ,\qquad \text{for } 
\alpha,\beta,\gamma \in \{1,2,3\} \; .
\end{align}
These complex structures on the cone are, of course, induced by the three Sasaki structures on the underlying 3-Sasakian manifold and 
give rise to a triplet of Kähler forms
\begin{align}
\label{eq:def_Kahler_forms}
 \Omega_{\a}(X,Y) \coloneqq g(X,J_{\a}(Y)) \; ,\qquad \text{for } 
\alpha \in \{1,2,3\} \; ,
\end{align}
whose components satisfy relations analogous to  \eqref{eq:quat_relations}.

\paragraph{Notation.}
For our discussion we apply the notations used in \cite{harland_noelle}. A 3-Sasakian manifold is then described by an orthonormal frame
of 1-forms  $e^1,  \ldots, e^{4m+3}$, where $e^{\a} \equiv \eta_{\a}$ for 
$\a=1,2,3$ are the duals of the characteristic vector fields $\xi_{\a}$, 
and 2-forms 
\begin{align}
\label{eq:3sasaki_2forms}
\begin{aligned}
\omega_1 &= \sum_{i=1}^{m} \left( e^{4i} \wedge e^{4i+1} 
+ e^{4i+2} \wedge e^{4i+3} \right) \, ,  \\
\omega_2 &= \sum_{i=1}^{m}\left( e^{4i} \wedge 
e^{4i+2} 
- e^{4i+1} \wedge e^{ 4i+3} \right)\, , \\
\omega_3 &= \sum_{i=1}^{m} \left( e^{4i} \wedge e^{4i+3} 
+ e^{4i+1} \wedge e^{4i+2} \right) \; ,
\end{aligned}
\end{align}
which are part of the exterior derivatives of $\eta^{\a}$ as follows:
\begin{align}
\label{eq:def_equ_3sasaki}
\diff \eta_{\a} = \epsilon_{\a}^{\ \b\g} \eta_{\b} \wedge\eta_{\g}  + 2 
\omega_{\a} 
  \quad \quad \lb \Longrightarrow \ \diff \omega_{\a}=2 \epsilon_{\a}^{\ \b\g} 
\eta_{\b} \wedge \omega_{\g}\rb.
\end{align}
The metric of the cone $g_{c}$ (or the conformally equivalent cylinder 
$g_{\mathrm{cyl}}$) reads
\begin{align}
g_{c}= r^2 \sum_{\m=1}^{4m+3} e^{\m} \otimes e^{\m} + \diff r \otimes 
\diff r \= r^2 \sum_{\m=0}^{4m+3} e^{\m} \otimes e^{\m} \equiv r^2 
g_{\mathrm{cyl}} \; ,
\end{align} 
with the definition $e^{0} \coloneqq \diff \tau \coloneqq \tfrac{\diff r}{r}$. 
The induced Kähler forms on the cone read
\begin{align}
\Omega_{\a} = r^2 (\omega_{\a} +\tfrac{1}{2} \epsilon_{\a\b\g} e^{\b \g} + 
\diff \tau \wedge e^{\a}).
\end{align}
We note that their closure follows from 
\eqref{eq:def_equ_3sasaki}. Explicitly, we have
\begin{align}
\label{eq:kaehler_forms}
\begin{aligned}
\Omega_{1}&= r^2\sum_{i=0}^{m} \left( e^{4i} \wedge e^{4i+1} 
+ e^{4i+2} \wedge e^{4i+3} \right) \,, \\
\Omega_{2}&=r^2 \sum_{i=0}^{m} \left( e^{4i} \wedge e^{4i+2} 
- e^{4i+1} \wedge e^{ 4i+3} \right) \,, \\
\Omega_{3}&=r^2\sum_{i=0}^{m}\left( e^{4i} \wedge e^{4i+3} 
+ e^{4i+1} \wedge e^{4i+2} \right) \, ,
\end{aligned}
\end{align}
where the summation now starts with $i=0$, in contrast to the expressions for 
$\omega_{\a}$ in
\eqref{eq:3sasaki_2forms}. On the tangent space
they induce the complex structures\footnote{They can also be obtained 
by writing quadruples $\mathbb{X}_i \coloneqq X_{4i} + \im X_{4i+1} + 
\mathrm{j} X_{4i+2} + \mathrm{k}X_{4i+3}$ and letting
$J_1=I$, $J_2=J$ and $J_3=K$ act on them by multiplication with $\im$, 
$\mathrm{j}$ and $\mathrm{k}$.} acting on basis vector fields $E_0,\ldots, 
E_{4m+3}$ as 
\begin{align}
\label{eq:action_j}
J_{\a} E_{4i} &= - E_{4i+\a} \quad \text{and} \quad J_{\a} E_{4i+\b} \= - 
\epsilon_{\a\b\g} E_{4i+\g}  \quad (\a \neq \b)
\end{align}
and similarly the action on the basis 1-forms reads
\begin{align}
J_{\a} e^{4i} &= e^{4i+\a} \quad \text{and} \quad J_{\a} e^{4i+\b} \= 
\epsilon_{\a\b\g} e^{4i+\g} \quad (\a \neq \b).
\end{align}
for $i=0, \ldots m$.
%
%
\section{Comments on moduli space of instantons}
\label{sec:comments_moduli_space}
Having established the notation, we proceed by a discussion of generic features 
for $\surm(n)$ and $\sprm(m)$-instantons. First, we consider the space of 
connections on hyper-Kähler spaces. Next, we provide the equivalent formulation 
of the $\sprm(m)$-instanton equations as intersection of three HYM instanton 
equations. Lastly, we introduce the ansatz for the connection on the cone (or 
conformally equivalent cylinder) over the Sasaki-Einstein or 3-Sasakian base, 
which reduces the instanton equations to Nahm-type equations.
\subsection{Space of connections over hyper-Kähler spaces}
In this section we describe the space of connections over a hyper-Kähler 
manifold $\mfd{4m}$ and show that it is equipped with a (formal) hyper-Kähler 
structure, which is induced from $\mfd{4m}$. This account is inspired from the 
analogous implication for the space of connections over Kähler manifolds, for 
which we refer to \cite{Atiyah:1982fa,Deser:2014zya,Sperling:2015sra}.
\paragraph{Preliminaries.}
Suppose $\mfd{4m}$ is a (closed) hyper-Kähler manifold of 
$\dim_{\HH}(\mfd{4m})=m$ and $G$ is a compact matrix group with 
$\gfrak=\mathrm{Lie}(G)$. We denote by $P(M^{4m},G)$ a $G$-principal bundle 
over $\mfd{4m}$, $\Int(P)\coloneqq P\times_G G$  the group bundle, 
$\Ad(P)\coloneqq P \times_G \gfrak $ the Lie algebra bundle, and $E\coloneqq 
P\times_G F$ an associated vector bundle (with vector space $F$ that carries a 
representation of $G$).

Then $\ca$ is a connection 1-form with curvature $\cf_\ca=\diff \ca + \ca 
\wedge \ca$, and $\Aa(P)$ (and $\Aa(E)$) denotes the space of connections on 
$P$ (and $E$).
The gauge group $\Ggauge$ can be identified with the global section on 
$\Int(P)$, i.e.\ 
\begin{align}
\label{eq:def_gauge_group}
\begin{aligned}
 \Ggauge &= \Gamma(\mfd{4m},\Int(P)) \; , \\
 \ca &\mapsto \ca^g \coloneqq \Ad(g^{-1}) \ca + g^{-1} \diff g \; , \quad 
\text{for } g \in \Ggauge  \; .
\end{aligned}
\end{align}
The associated Lie algebra $\ggauge$ of $\Ggauge$ is identified with the global 
section on $\Ad(P)$, i.e.\
\begin{align}
\label{eq:def_gauge_algebra}
\begin{aligned}
 \ggauge &= \Gamma(\mfd{4m},\Ad(P)) \; , \\
 \ca &\mapsto \delta\ca = \diff_\ca \chi \coloneqq \diff \chi  + [ \ca, 
\chi] \; , \quad \text{for } \chi \in \ggauge  \; .
\end{aligned}
\end{align}
Moreover, $\Aa(P)$ is an affine space over $\Omega^1(\mfd{4m},\Ad(P))$; thus, 
the tangent space $T_\ca \Aa$ for any $\ca \in \Aa(P)$ can be canonically 
identified with $\Omega^1(\mfd{4m},\Ad(P))$. By assumption, $G \hookrightarrow 
\urm(N)$, for some $N\in \N$; thus, the trace provides an $\Ad$-invariant inner 
product.
\paragraph{Metric.}
A Riemannian structure on $\Aa(P)$ is established via
\begin{align}
\label{eq:def_metric_connection}
 \bg_{|\ca}(X,Y)\coloneqq \int_{\mfd{4m}} \tr \left( X \wedge \star Y \right) 
\; , \quad \text{for } X,Y \in T_\ca \Aa \; ,
\end{align}
which is symmetric and base-point independent. Moreover, the definition 
employs the metric structure on the base manifold via the Hodge star $\star$.
\paragraph{Symplectic forms.}
Similarly, one can define three symplectic structures on $\Aa(P)$ via
 \begin{align}
 \label{eq:def_symplectic_connection}
 (\bw_\alpha)_{|\ca}(X,Y)\coloneqq \int_{\mfd{4m}} \tr \left( X \wedge Y 
\right) \wedge \frac{(\Omega_\alpha)^{2m-1}}{(2m-1)!}
\; , \quad \text{for } X,Y \in T_\ca \Aa \; , \alpha=1,2,3 \;,
\end{align}
which is skew-symmetric and base-point independent. Again, the entire 
$\C P^1$ -worth of symplectic structures of the base manifold transfers to a 
$\C P^1$  of symplectic structures on $\Aa(P)$. To show that $\bw_\alpha$ is 
non-degenerate one can explicitly verify that 
\begin{align}
 \star J_\alpha(Y) = Y \wedge \frac{(\Omega_\alpha)^{2m-1}}{(2m-1)!} \; 
, \quad \forall \alpha =1,2,3 \; .
\end{align}
Here $J_\alpha$ acts only on the $1$-form part of $Y$. Consequently,
 \begin{align}
 (\bw_\alpha)_{|\ca}(X,Y)= \int_{\mfd{4m}} \tr \left( X \wedge \star 
J_{\alpha}(Y) 
\right)  = \bg_{|\ca}(X,J_\alpha(Y))
\; , \quad \text{for } X,Y \in T_\ca \Aa \; , \alpha=1,2,3 \;,
\end{align}
and $\bw_\alpha$ is non-degenerate because $\bg$ is.
\paragraph{Complex structure.}
Having a Riemannian and three symplectic structures on $\Aa(P)$ it is tempting 
to introduce the compatible complex structures $\bJ_\alpha$ via
\begin{align}
 \bw_\alpha(\cdot, \cdot) = \bg(\cdot,\bJ_\alpha(\cdot)) \; .
\end{align}
It follows from the above that\footnote{Compared to \cite{Sperling:2015sra}, we 
consider $\bJ= -\bJ_{\text{can}}$, where $\bJ_{\text{can}}$ is the 
canonical complex structure defined via $\bw(\cdot, \cdot) = 
\bg(\bJ_{\text{can}}(\cdot),\cdot)$ on $\Aa(P)$.}
\begin{align}
\label{eq:def_cplx_str_connection}
 \bJ_\alpha (Y) = J_\alpha(Y) \; , \quad \text{for } Y \in T_\ca \Aa \; , 
\alpha=1,2,3 \;.
\end{align}
Thus, the three complex structures on $\Aa(P)$ are base-point independent, are 
induced from the complex structures on $\mfd{4m}$, and, consequently, satisfy 
the quaternionic algebra.

In summary, $\Aa(P)$ (and also $\Aa(E)$) is equipped with a (formal) 
hyper-Kähler 
structure, inherited from $\mfd{4m}$, and a compatible $\Ggauge$-action. We 
will see in a moment that the moduli space of the hyper-Kähler instanton 
equations can be understood as a hyper-Kähler quotient thereof.
\subsection{Equivalence of \texorpdfstring{$\sprm(m)$}{Sp(m)}-instantons and 
intersections of HYM instantons}
\label{sec:equivalence_Sp_HYM}
Let $\mfd{4m}$ be a hyper-Kähler manifold with complex structures 
$J_\alpha$ for $\alpha=1,2,3$ satisfying \eqref{eq:quat_relations} and 
corresponding Kähler forms $\Omega_\alpha$, defined via 
\eqref{eq:def_Kahler_forms}. 
One can parametrise a $\C P^1$ of complex structures via $s_\alpha \in \R$, 
$\delta^{\alpha \beta} s_\alpha s_\beta=1$ such that any complex structure (and 
corresponding Kähler form) can be written as 
\begin{align}
 J \coloneqq s^\alpha J_\alpha \; , \qquad \Omega \coloneqq s^\alpha 
\Omega_\alpha \; . 
\end{align}

Consider a connection $\ca$ on a complex vector bundle $E$ over $\mfd{4m}$. 
Since $\mfd{4m}$ is hyper-Kähler the generic holonomy 
$\sorm(4m)$ is reduced to $\sprm(m)$, and one has the splitting 
\begin{align}
 \sormL(4m) = \sprmL(m) \oplus \sprmL(1) \oplus \mathfrak{k}.
\end{align}
Following the definition of instantons on $G$-structure 
manifolds\footnote{Equivalently, one can define $\sprm(m)$-instantons in terms
of a generalised self-duality condition; for details, see \cite{harland_noelle}.}, 
 $\sprm(m)$ instantons are defined as connections such that the curvature $2$-form $\cf_\ca$ takes values in the Lie algebra 
$\sprmL(m)$ only, i.e.\ the instanton equations are equivalent to the 
vanishing of the $\sprmL(1)\oplus \mathfrak{k}$-part of the curvature $2$-form. 
 
According to \cite{Salamon:1988,Nitta:1988}, the $\sprm(m)$-instanton equations 
can be recast as 
\begin{align}
 \cf_{J}^{0,2} =0 \quad \text{ for all $J$,}
 \label{eq:tri_holo}
\end{align}
i.e.\ they can be obtained from the holomorphicity conditions for any complex 
structure $J$. Recall that for a fixed $J=J_{\a}$ the
holomorphicity condition only  induces the reduction of the holonomy algebra as
\begin{align}
 \sormL(4m) = \urmL_\alpha(2m) \oplus \mathfrak{P}_\alpha \; ,
\end{align}
while HYM instantons additionally constrain the  $\mathfrak{u}_{\a}(1)$ part 
of the splitting $\urmL_\alpha(2m) = \mathfrak{u}_{\a}(1) \oplus  \mathfrak{su}_{\a}(2m)$ by imposing the stability-like condition
\begin{align}
 \Omega_\alpha^{\mu \nu} \cf_{\mu \nu} =0 \; , \quad \alpha =1,2,3 \; .
 \label{eq:stab}
\end{align}
However, satisfying the holomorphicity condition \eqref{eq:tri_holo} for \emph{any} $J$ already implies the stability-like conditions,
as it is shown e.g.\ in Section 4.5 of \cite{s7}. It can be also seen in the 
explicit instanton equations we use for the discussion in Section 
\ref{sec:explicit_inst_eq}.  

Hence, it is justified to consider the 
moduli space of $\sprm(m)$-instantons as the intersection
\begin{align}
\label{eq:intersection_mod_space}
 \cm_{\sprm(m)} = \cm_{\surm_1(2m)}\cap  
\cm_{\surm_2(2m)}\cap
\cm_{\surm_3(2m)} = \bigcap_{J} \cm_{\surm_J(2m)} \; .
\end{align}
\subsection{Quaternionic instantons}
As shown in the previous section, we can understand the
$\sprm(m)$-instanton conditions on $E \to \mfd{4m}$ as
\begin{align}
\label{eq:inst_triple_HYM}
 \cf_\alpha^{(0,2)} =0 \qquad \text{and} \qquad 
 \Omega_\alpha \lrcorner \cf_\alpha =0 \qquad \forall \alpha=1,2,3 \; .
\end{align}
As explained, for instance, in \cite{Sperling:2015sra}, the condition 
$\cf_\alpha^{(0,2)} =0$ introduces a holomorphic structure on the vector bundle 
$E$. Since we have three holomorphic structures arising, the bundle 
becomes tri-holomorphic. Denote the space of tri-holomorphic connections as
\begin{align}
 \Aa^{\mathrm{holo}}(E) = \left\{ \ca \in \Aa(E) | \cf_\alpha^{(0,2)} =0 \;, 
\forall \alpha=1,2,3 \right\} \; .
\end{align}
We expect that $\Aa^{\mathrm{holo}}(E)$ is equipped with a hyper-Kähler 
structure by restriction from $\Aa(E)$ and has a compatible action of $\Ggauge$.

On $\Aa^{\mathrm{holo}}(E) $, the three remaining equations $\Omega_\alpha 
\lrcorner \cf_\alpha =0$ are understood as triplet of moment maps $\mu_\alpha$ 
for the gauge group. The proof of the statement is a generalisation of 
\cite{Atiyah:1982fa} and has been shown in \cite{Sperling:2015sra} for the Kähler 
case. Since the arguments are identical, we refrain from repeating them here.

It is, however, important to realise that the case of non-compact hyper-Kähler 
cones requires one to consider the \emph{framed} gauge group $\Ggauge_0$ for 
the moment maps to be well-defined.

Thus, we presume that  the  moduli space of hyper-K\"ahler instantons can be expressed as 
(trivial) hyper-Kähler 
quotient
\begin{align}
 \cm_{\sprm(m)} = \left\{ \ca \in\Aa^{\mathrm{holo}}(E) \ \big| \ \mu_\alpha=0  
\; 
,\forall \alpha =1,2,3 \right\} \slash \Ggauge =  \Aa^{\mathrm{holo}}(E)  \slash 
\Ggauge \; .
\end{align}
The arguments presented earlier imply that it is a trivial 
quotient in the sense that the moment map 
conditions are already satisfied on all of $\Aa^{\mathrm{holo}}(E) $. However, 
the consequence remains true; $\cm_{\sprm(m)} $ is itself a hyper-Kähler space.
%
%
\subsection{Ansatz for equivariant instantons}
\label{sec:Ansatz_hk}
Before we investigate the instantons on metric cones we briefly 
describe the set-up, which is based on the approach of 
\cite{ivanova_popov} and has been thoroughly discussed in 
\cite{Bunk:2014coa,Sperling:2015sra}.

Consider $H=\surm(n)$ or $\sprm(m)$ as closed subgroup of 
$G=\surm(n{+}1)$ or $\sprm(m{+}1)$, respectively.  
Let $\mfd{k}$  ($k=2n{+}1$ for Sasaki-Einstein and $k=4m{+}3$ 
for 3-Sasakian) be a manifold with
$G$-structure together with a canonical connection $\Gamma^P$ on the 
tangent bundle, see \cite{harland_noelle}. The metric cone, by choice of our 
examples, is a manifold with reduced holomony $G\subset \sorm(k+1)$.
By conformal invariance of the instanton equations, 
we can equally well consider $\mathrm{Cyl}(\mfd{k})$, which is equipped with 
a non-integrable $G$-structure. Let $P$ be the principal 
$G$-bundle of the frame bundle of $\mathrm{Cyl}(\mfd{k})$ which 
comprises this $G$-structure and associate a complex vector 
bundle $E \to \mathrm{Cyl}(\mfd{k})$ of rank $p$. 
The fibres $E_x \cong \C^p$ are equipped with a Hermitian form.

Thus, the connection 1-from associated to any $\ca$ is a 
$\gfrak$-valued $1$-form on $\mathrm{Cyl}(\mfd{k})$. We consider an 
ansatz of the form
\begin{subequations}
\label{eq:ansatz_gauge_conn}
\begin{align}
 \ca = \widehat{\Gamma}^P + X \; ,
\end{align}
with $\widehat{\Gamma}^P$ denoting the lifted $\hfrak$-valued connection on 
$E$ obtained from $\Gamma^P$. On a patch $\mathcal{U} \subset 
\mathrm{Cyl}(\mfd{k})$ with a basis of $1$-forms 
$(e^0,\{e^\mu\}_{\mu=1}^{k})$ we 
can describe $X$ via
\begin{align}
X_{|\mathcal{U}} = X_{0}\otimes e^0 + X_\mu \otimes e^\mu \; ,
\end{align}
\end{subequations}
with ${X_\mu}_{| x} \in \End(\C^p)$ for $x\in \mathcal{U}$.
It is customary to eliminate $X_0$ by a suitable gauge transformation --- 
called \emph{temporal gauge} --- but there is no need to do this.

So far, this is just a particular way of rewriting a generic connection. 
However, we further restrict to connections for which the endomorphisms-valued 
functions $X_\mu$, firstly, 
depend only on the cone / cylinder coordinate, and, secondly, satisfy an 
\emph{equivariance condition}. Since $H$ is a closed subgroup of 
$G$ one has 
the $H$-invariant decomposition
\begin{align}
 \mathrm{span} \langle I_{M}\rangle \equiv 
 \gfrak =  \hfrak \oplus \mathfrak{m}
 \equiv \mathrm{span} \langle I_j\rangle 
 \oplus \mathrm{span} \langle I_{\m}\rangle. 
 \label{eq:decomp_algebra}
\end{align}
Denote by $\widehat{I}_M$ the generators in the representation on the fibres of 
$E_x \cong \C^{p}$. The generators satisfy the following commutation relations:
\begin{align}
 \com{\widehat{I}_j}{\widehat{I}_k} = f_{jk}^{\ \ l} \widehat{I}_l \; , \qquad
 \com{\widehat{I}_j}{\widehat{I}_\m} = f_{j\m}^{\ \ \n} \widehat{I}_\n \; , 
\qquad
 \com{\widehat{I}_\m}{\widehat{I}_\n} = f_{\m \n}^{\ \ j} \widehat{I}_j 
 +f_{\m \n}^{\ \ \sigma} \widehat{I}_\sigma\; .
 \label{eq:algebra}
\end{align}
Then the equivariance conditions read\cite{ivanova_popov}
\begin{align}
\label{eq:def_equivariance}
 \com{\widehat{I}_j}{X_\mu} = f_{j \mu}^{\ \ \nu} X_\nu \; , \qquad
 \com{\widehat{I}_j}{X_0} = 0 \; .
\end{align}
These conditions can be satisfied, for instance, by choosing the matrix-valued 
functions $X_\mu$ proportional to the generators spanning $\mfrak$, i.e.\ 
$X_{\m}=\l_{\m}(r) \widehat{I}_{\m}$, so that the instanton equations reduce 
to equations on the scalar functions $\l_{\m}(r)$ only. This approach has been
pursued in the constructions of instantons in various settings, see for 
instance \cite{ivanova_popov,Bunk:2014kva,Bunk:2014coa,Haupt:2015wdq}. Solving 
the equivariance condition more generally leads to quiver gauge theories 
\cite{Lechtenfeld:2008nh,Dolan:2010ur,Lechtenfeld:2015ona,
Geipel:2016uij,Geipel:2016hpk,s7} that depend on the chosen manifold.
For the moment, we \emph{suppose} that one has implemented the 
equivariance conditions and is left  with the relevant instanton equations.
 We comment on the equivariance 
condition in Section \ref{sec:equivariance}.
As a remark, not imposing \eqref{eq:def_equivariance} amounts to dimensional 
reduction instead of an equivariant reduction, which is legitimate by itself.

In summary, we search for connections satisfying \eqref{eq:def_equivariance} 
and the instanton equations simultaneously.
For this ansatz, the gauge group \eqref{eq:def_gauge_group} reduces 
to\footnote{Including the equivariance at this stage would imply the 
decomposition of the gauge group $\{ g : \R \to \prod_k 
\urm(V_k)\}$ following the decomposition of the endomorphisms space 
$\End(\C^p)|_{H}=\oplus_k V_k$ on the typical fibre $E_x\cong \C^p$.}
\begin{align}
 \Ggauge = \{g : \R \to \urm(p) \} \; ,
 \label{eq:gauge_group_cone}
\end{align}
which acts on the matrix-valued functions as follows:
\begin{align}
 X_\mu \mapsto X_\mu^g \coloneqq \Ad(g) X_\mu 
 \;, \qquad 
 X_0 \mapsto X_0^g 
\coloneqq \Ad(g) X_0 - \frac{1}{2} \left( \frac{\diff }{\diff t} g\right) 
g^{-1} \; .
\label{eq:gauge_trafo}
\end{align}
As mentioned earlier, due to the non-compactness of the metric cone we need to 
restrict ourself to the framed gauge transformations.
%
%
%
\section{Instantons on Calabi-Yau cones}
\label{sec:instanton_cy-cone}
In this section we firstly recap the choice of boundary conditions used 
in \cite{Sperling:2015sra} and secondly introduce a different class of boundary 
conditions. This allows to parallel the HYM matrix instanton equations on the 
Calabi-Yau cone with the two choices of boundary conditions for Nahm's 
equations treated by Kronheimer in 
\cite{kronheimer1990hyper,Kronheimer:1990ay}. 
\subsection{Set-up}
Before exploring the details, we need to recall the set-up of Nahm's equations 
and the generalised Nahm's equations for Calabi-Yau instantons.
\paragraph{Nahm's equations.}
As customary, one splits Nahm's equations in a \emph{complex equation} 
\cite{donaldson1984}
\begin{subequations}
\label{eq:Nahm_eq}
\begin{equation}
 \frac{\diff \beta}{\diff t} + 2 \beta + 2\com{\alpha}{\beta} =0 
\end{equation}
and a \emph{real equation}
\begin{equation}
 \frac{\diff}{\diff t} (\alpha + \alpha^*) + 2(\alpha + \alpha^*) 
 + 2 \left( \com{\alpha}{\alpha^*} + \com{\beta}{\beta^*} \right) =0 \;,
\end{equation}
\end{subequations}
for $\alpha = \frac{1}{2} (A_0 + i A_1)$ and $\beta = \frac{1}{2} (A_2 + i 
A_3)$. The $A_j$ are the components of a connection on a $G$-bundle $P \to S^3 
\times \R$. The ``model'' solution, in temporal gauge $A_0=0$, is given by 
\cite{kronheimer1990hyper}
\begin{align}
 A_j =  e^{-2 t} \tau_j + \sigma_j
\end{align}
where $\tau_j$ are elements of a Cartan subalgebra of $\gfrak$ and $\sigma_j$ 
are elements of $\gfrak$ that commute with the $\tau_j$ and which satisfy the 
$\surmL(2)$ relations. In more detail, the $\sigma_j$ are critical points of a 
gradient flow; hence, they establish a Lie algebra homomorphism $\rho:\surmL(2) \to 
\gfrak$.

Kronheimer considers the two extreme cases: only $\tau_j$ in 
\cite{kronheimer1990hyper} and only $\sigma_j$ in 
\cite{Kronheimer:1990ay}. In both cases, the objective has been to establish 
the hyper-Kähler structure of certain coadjoint orbits of complex Lie groups 
via the known hyper-Kähler structure of the moduli space of Nahm's equation. 
The crucial point in the suitable identification lies in the choice of boundary 
conditions.

From the physics point of view\footnote{We refer to 
\cite[Sec.\ 3]{Gaiotto:2008sa} for an accessible review.}, the boundary 
conditions of \cite{Kronheimer:1990ay} 
\begin{equation}
 \lim_{t\to \infty} A(t) =0 \; , \qquad \lim_{t \to - \infty} A(t) \in C(\rho)
 \label{eq:bc_Nahm_instanton}
\end{equation}
are most interesting as they realise the correspondence between the 
instanton moduli space and nilpotent orbits of the complex Lie group 
$G^\C$. Whereas the regular boundary conditions of \cite{kronheimer1990hyper}, 
led to an identification of the moduli space with the maximal semi-simple 
orbit. 
\paragraph{Matrix instanton equations on Calabi-Yau cone.}
In temporal gauge, the instanton matrix equations considered in
\cite{Sperling:2015sra}, and also in \cite{ivanova_popov}, read\footnote{We keep 
the notation of \cite{Sperling:2015sra} and label the contact direction of the 
Sasaki-Einstein structure with $\eta = e^{2n+1}$.}
\begin{subequations}
\label{eq:CY_instanton}
\begin{alignat}{2}
 \com{X_{2j-1}}{X_{2k-1}} &=  \com{X_{2j}}{X_{2k}}   , &
 \com{X_{2j-1}}{X_{2k}} &= -  \com{X_{2j}}{X_{2k-1}}   ,
 \label{eq:CY_alg}\\
  \tfrac{\diff}{\diff t} X_{2j-1} + \tfrac{n{+}1}{n} 
X_{2j-1}&= \com{X_{2j}}{X_{2n+1}}  ,  \qquad&
   \tfrac{\diff}{\diff t} X_{2j} + \tfrac{n{+}1}{n} 
X_{2j} &= -\com{X_{2j-1}}{X_{2n+1}} ,
\label{eq:CY_diff}
\end{alignat}
for $j,k=1,\ldots, n$ and
\begin{align}
 \tfrac{\diff }{\diff t} X_{2n+1} + 2n X_{2n+1} = \sum_{k=1}^{n} 
\com{X_{2k-1}}{X_{2k}} \; .
\end{align} 
\end{subequations}
The novel insight, compared to \cite{ivanova_popov,Sperling:2015sra}, is that 
the appearing matrix differential equations can be written as gradient flow 
$\frac{\diff }{\diff t} X = - \nabla \Psi(X)$ for
\begin{align}
 \Psi(X_\mu) \coloneqq \frac{n+1}{2n} \sum_{a=1}^{2n} \tr(X_a X_a) 
 + n \ \tr(X_{2n+1} X_{2n+1}) 
 - \tr\left(X_{2n+1}   \sum_{k=1}^{n} \com{X_{2k-1}}{X_{2k}}  \right) \; ,
 \end{align}
while the algebraic conditions \eqref{eq:CY_alg} have to be imposed as 
additional constraints. Nevertheless, the additional constraints are preserved 
by the flow; hence, they only need to hold at one $t_0\in \R$ in order to hold 
at any other instance.

This gradient flow formulation is a reflection of the known 
phenomenon\cite{Haupt:2011mg,harland_noelle,Lechtenfeld:2012yw} that 
the generalised instanton equations, in temporal gauge, on a cylinder over a 
manifold $M$ are equivalent to the generalised Chern-Simons gradient flow on 
$M$ subject to additional constraints. For special cases, like $3$-manifolds or 
7-manifolds with nearly parallel $G_2$ structure, the additional constraints 
are implied by the gradient flow.

The generic model solution for \eqref{eq:CY_instanton} is of the 
form, see also \cite{Sperling:2016dqk},
\begin{align}
 X_a=e^{-\frac{n+1}{n}t} T_a + S_a \; ,\;  a=1,\ldots,2n \, , \qquad
 X_{2n+1}=e^{-2nt} T_{2n+1} + S_{2n+1} \;,
\end{align}
where the $T_\mu$, for $\{\mu\} = \{a, 2n+1\}$, lie in a Cartan subalgebra, 
$\com{T_\mu}{S_\nu}=0$ for all $\mu ,\nu$, and the $S_\mu$ are critical points 
of $\Psi$ subject to the algebraic conditions \eqref{eq:CY_alg}.

This can be put in context to the treatment of Nahm's equations:
Firstly, the regular boundary conditions for the HYM instantons on Calabi-Yau 
cones of \cite{Sperling:2015sra} will lead to a diagonal orbit in which 
the moduli space can be embedded. 
Secondly, boundary conditions  similar to \cite{Kronheimer:1990ay} for 
the HYM instanton equations have not yet been considered. 
For the Calabi-Yau instantons it is not straightforward to adapt Kronheimer's 
analysis, because the critical points of $\Psi$, even imposing the additional 
constraints \eqref{eq:CY_alg}, do not necessarily give rise to a Lie algebra 
homomorphism.
Nevertheless, one could study boundary conditions for which the 
$S_\mu$ define a Lie algebra homomorphism of $\surmL(n{+}1)$ in $\urmL(p)$. 
This will be the subject of  a later section.

For most of the analysis of the next two sections one only requires the form of
the complex equations \eqref{eq:CY_alg}--\eqref{eq:CY_diff}.
We can rewrite the \emph{complex equations} in the complexified basis and 
find
\begin{align}
\label{eq:cplx_eq_CY}
\begin{aligned}
 \frac{\diff}{\diff t} Y_j + \frac{n+1}{n} Y_j + 2 \com{Y_{n+1}}{Y_j} &=0 \, ,\\
 \com{Y_j}{Y_k} &=0 \, ,
 \end{aligned}
\end{align}
for $j,k=1,\ldots,n$.
Moreover, the linear terms in the instanton matrix equations can be eliminated 
by a 
suitable rescaling. For the rescaled matrices we use the notation
\begin{align}
 X_a = e^{-\frac{n+1}{n}t} \Xcal_a \; , \qquad  
 X_{2n+1} = e^{-2nt} \Xcal_{2n+1} \; , \qquad  
 X_{2n+2} = e^{-2nt} \Xcal_{2n+2} \;,
\end{align}
and analogously for $Y_j \mapsto \Ycal_j $.
This rescaling is accompanied by a new variable $  s\coloneqq -\frac{1}{2n} 
e^{-2nt}\in \R^-$.
\subsection{Relation to coadjoint orbits}
The boundary conditions considered in \cite{Sperling:2015sra} are
\begin{equation}
\label{eqn:boundary_cond}
 \exists\,  g_0 \in \urm(p) \text{  such that } \forall 
\mu=1,\ldots,2n+1: \lim_{s\to 
-\infty}\Xcal_{\mu}(s) =\Ad(g_0) \, T_{\mu} \;,
\end{equation}
where the $T_{\mu}$ lie in a Cartan subalgebra of $\surmL(p)$.
For simplification, we can \emph{require} the $T_\mu$ to be a \emph{regular 
tuple}, i.e.\ the intersection of the centralisers of the $T_\mu$ 
consists only of the Cartan subalgebra of $\surmL(p)$. Then all of the $S_\mu$ 
have to vanish such that the $T_\mu$ alone provide the only model for the 
behaviour of the $\Xcal_\mu$ near $s\to -\infty$.

Let us denote by $\Mcal_n(E)$  the moduli space of solutions to the 
complex and real equations satisfying the boundary 
conditions~\eqref{eqn:boundary_cond} (with suitable regularity) as well as the 
equivariance condition.
From the considerations presented in \cite{Sperling:2015sra}, we can establish 
the following map
\begin{equation}
\label{eqn:map}
 \begin{split}
  \Mcal_n(E) &\to \mathcal{O}_{\mathrm{diag}}(\Ycal_1,\ldots,\Ycal_n) \\
(\Ycal,\Zcal) &\mapsto (\Ycal_1(0),\ldots,\Ycal_n(0))
 \end{split}
\end{equation}
where $\mathcal{O}_{\mathrm{diag}}(\Ycal_1,\ldots,\Ycal_n)$ is defined as 
follows:
The $n$ commuting objects $\Ycal_k$ can be understood as element of 
$\glrmL(p,\C) \otimes 
\C^n$, because the gauge group $ \glrm(p,\C)$ does not act separately on each 
$\Ycal_k$, but it acts the \emph{same} on every $\Ycal_k$. In other words, 
consider $\left(\glrm(p,\C)\right)^{\times n}$  with the diagonal embedding 
$\glrm(p,\C) \hookrightarrow  \glrm(p,\C)^{\times n}$, which gives rise to the 
relevant action~\eqref{eq:gauge_group_cone}. Then we see
\begin{equation}
\label{eq:def_adj_orbit_diag}
\begin{aligned}
 \mathcal{O}_{\mathrm{diag} }(\Ycal_1,\ldots,\Ycal_n) &\coloneqq 
 \left\{ \left(\Ad(g)\Ycal_1(0), \ldots \Ad(g)\Ycal_n(0)\right) \, \big| \, g 
\in 
\glrm(p,\C) \right\} \\
 &\subset \prod_{j=1 }^n \left\{ \Ad(g_j) \Ycal_j(0) \, \big| \, g_j \in 
\glrm(p,\C) \right\} = 
\mathcal{O}_{\Tcal_1} \times \cdots \times \mathcal{O}_{\Tcal_n}  
\end{aligned}
\end{equation}
where $\mathcal{O}_{\Tcal_k} $ denotes the adjoint orbit of $\Tcal_k$  in 
$\glrmL(p,\C)$.  
Analogous to~\cite{kronheimer1990hyper}, the map~\eqref{eqn:map} is injective 
due 
to the  uniqueness of the corresponding solution of the real and complex 
equations. In contrast, the surjectivity is less clear. By the construction of 
the local solution \cite[Eq.\ (3.40)]{Sperling:2015sra}, one finds that any 
element of 
$\mathcal{O}_{\mathrm{diag} }(\Ycal_1,\ldots,\Ycal_n)$ gives rise to a solution 
of the complex and real equation, but it is unclear if this solution satisfies 
the required asymptotic. 

Moreover, one knows that the orbit of an element $\Tcal_k$ of the Cartan 
subalgebra is of the form $\glrm(p,\C)/ \mathrm{Stab}(\Tcal_k) $, where 
$\mathrm{Stab}(\Tcal_k) $ is the maximal torus of $\glrm(p,\C)$ because each 
$\Tcal_k$ is assumed to be a regular element. 
The product of the regular semi-simple coadjoint orbits is a complex symplectic 
manifold. Each orbit is equipped with the  Kirillov-Kostant-Souriau 
symplectic form and the product thereof gives the symplectic structure on the 
total space. 
As a manifold the orbit $\mathcal{O}_{\mathrm{diag} 
}(\Ycal_1,\ldots,\Ycal_n)$ is just 
$\glrm(p,\C)\slash \mathrm{Stab}\left(\Ycal_1(0),\ldots,\Ycal_n(0)\right) $,
wherein
\begin{equation}
 \mathrm{Stab}\left(\Ycal_1(0),\ldots,\Ycal_n(0) \right) = \bigcap_{j=1}^{n} 
\mathrm{Stab}\left(\Ycal_j(0)\right) =  \bigcap_{j=1}^{n} 
\mathrm{Stab}(\Tcal_j) 
\end{equation}
and the intersection of the stabilisers of the $\Tcal_j$ is the 
complexified maximal torus, by the regularity assumption. Hence, the complex 
dimension\footnote{In fact, as each $\Tcal_j$ is a 
regular pair, each regular semi-simple $\mathcal{O}_{\Tcal_k}$ has the same 
dimension as the diagonal orbit.} is
\begin{equation}
 \dim_{\C} \left(\mathcal{O}_{\mathrm{diag} }(\Ycal_1,\ldots,\Ycal_n) \right)= 
 \dim_{\R}(\urm(p)) -\rank (\urm(p))  = p(p-1)\; ,
\end{equation}
which always is a multiple of $2$. The diagonal orbit is also Kähler, as it is 
a complex sub-manifold of a (hyper-)Kähler product.
Analogous to~\cite{kronheimer1990hyper}, the map~\eqref{eqn:map} is holomorphic 
such that it describes an embedding of 
the framed moduli space $\Mcal_n(E)$ into the diagonal orbit, which is a 
finite-dimensional Kähler manifold.
\subsection{Singular boundary conditions}
In addition, we can consider the other extreme case, in which the boundary 
conditions are determined by the critical points of the gradient flow.

As mentioned earlier, the equations determining a critical point are not 
sufficient to define a Lie algebra homomorphism, but they are compatible with a 
Lie algebra homomorphism. Moreover, note that the basis elements of $\mfrak$ in 
the decomposition \eqref{eq:decomp_algebra}
are sufficient to generate $\surmL(n{+}1)$ as algebra, see also 
Appendix \ref{app:details}.

Inspired by the boundary conditions chosen in \cite{Kronheimer:1990ay}, suppose 
we have two Lie algebra homomorphisms $\rho_-$ and $\rho_+$. Then we consider 
boundary conditions of the type\footnote{The precise formulation is given in 
\eqref{eq:def_cplx_trajectory}.}
\begin{align}
\lim_{t\to - \infty} X(t) \in C(\rho_-) \; , \qquad 
\lim_{t\to + \infty} X(t) = \rho_+ \; .
\label{eq:bc_matrics_X}
\end{align}
Here, $C(\rho_-)$ consists of all homomorphisms conjugated to $\rho_-$ under 
the adjoint action of $\urm(p)$.

The treatment of the instanton matrix equations is as in the case of Nahm's 
equations: firstly, consider the complex equations with the boundary conditions 
and identify the equivalence classes of complex trajectories. Secondly, show 
that each solution of the complex equations can be gauge transformed into a 
solution of the real equation and that this gauge transformation is unique.

We delegate the details of the ``complex trajectories'', i.e.\ solutions to the 
complex equations satisfying the appropriate boundary conditions, to Appendix 
\ref{app:details}, while the treatment of the real equation can be taken over 
from \cite{Sperling:2015sra}. We find that the equivalence classes of the 
complex trajectories associated to 
homomorphisms $\rho_{\pm}$ are parametrised by a nilpotent orbit and, abusing 
the name, a ``transverse slice'' as
\begin{align}
 \mathcal{N}_{\diag}(\rho_-) \cap S_{\diag}(\rho_+) 
\end{align}
with
\begin{align}
 \Ncal_{\diag}(\rho_-) &\coloneqq
 \left\{ (\xi_1,\ldots,\xi_n) \in \glrmL(p,\C) \otimes \C^n
 |   
 \Ad_g (\xi_1,\ldots,\xi_n) = (F_1^-,\ldots,F_n^-) \; , g \in \glrm(p,\C)
\right\} \notag \\
&\subset \Ncal(F_1^-) \times \ldots \times \Ncal(F_n^-)
\label{eq:def_nilpotent_orbit_diag}
\end{align}
where $\Ncal(F_j^-)$ is the $\glrm(p,\C)$ nilpotent orbit of $F_j^- \coloneqq \rho_- (F_j)$, with $F_j$ defined in 
\eqref{eq:def_nilpotent_elements}.
Also, we have defined
\begin{align}
 S_{\diag}(\rho_+) &\coloneqq (F_1^+,\ldots,F_n^+) + 
 z(E_1^+) \times \ldots \times z(E_n^+)
\subset (\glrmL(p,\C))^{\times n} \,.
\label{eq:def_transvers_slice_diag}
\end{align}
Compared to the corresponding expressions in \cite{Kronheimer:1990ay} the orbit of the (unique) nilpotent element $Y$ in the case
of $\mathrm{SU}(2)$ had to be replaced by a diagonal\footnote{It has to be 
diagonal due to the gauge transformations \eqref{eq:gauge_trafo} that act 
with the same group element on all $\Ycal_j$.} orbit 
of an $n$-tuple of commuting nilpotent elements $F_j$. 
One can assign a notion of nilpotency to this diagonal orbit either naively 
in the sense that for $(\xi_1,\ldots,\xi_n)\in  
\Ncal_{\diag}(\rho_-)$
\begin{align}
 (\xi_1,\ldots,\xi_n)^k \equiv  (\xi_1^k,\ldots,\xi_n^k)= (0,\ldots, 0) 
\end{align}
for sufficiently large $k$ since each $\xi_j$ is nilpotent, or on more 
general grounds in the context of nilpotent pairs, which we comment on 
below in Section \ref{sec:generalised_Nahm}.
Similarly, the expression for the ``transverse slice'' $\mathcal{S}$ is adapted 
by considering $n$-tuples of nilpotent elements and the centralisers of the 
elements of $E_j^+ \coloneqq \rho_+(E_j)$ that generalise the matrix $X$ in 
Kronheimer's discussion.

Following \cite[App.\ A.4]{Sperling:2015sra}, the treatment of the real 
equation reduces to two statements: (i) for every complex trajectory there exists a 
gauge transformation such that the real equation holds, and (ii) equivalent 
complex trajectories, both satisfying the real equation, are related by a gauge 
transformation. Denote by $\widetilde{\Mcal}_n(E)$ the  space of solutions to 
the instanton matrix equations satisfying \eqref{eq:bc_matrics_X}, then we 
obtain the  map
\begin{align}
\begin{aligned}
 \widetilde{\Mcal}_n(E) &\to  \mathcal{N}_{\diag}(\rho_-) \cap 
S_{\diag}(\rho_+)  \\
(Y_j,Y_{n+1}) &\mapsto (Y_1(0),\ldots,Y_n(0)) \;,
\end{aligned}
\end{align}
which is clearly injective due to existence and uniqueness of the gauge 
transformation that renders a complex trajectory into a solution of the real 
equation. Again, surjectivity is not clear.

If the representation $\rho_+$ is the trivial representation, then $ 
S_{\mathrm{diag}}(\rho_+)$ is all of $(\glrmL(p,\C))^{\times n}$ such that the 
moduli space 
coincides with the diagonal orbit $\mathcal{N}_{\mathrm{diag}}(\rho_-)$. Note 
that this implies that the connection $\ca$ reduces to the (lifted) canonical 
connection $\Gamma^P$ as $t\to +\infty$. 
Consequently, the analogous boundary conditions to 
\eqref{eq:bc_Nahm_instanton}, i.e.\ 
\begin{align}
 \lim_{t\to -\infty} X(t) \in C(\rho_-) \;, \qquad 
 \lim_{t \to \infty } X(t) =0 \;,
\end{align}
lead to an embedding of $\widetilde{\Mcal}_n(E)$ into (the closure of) a 
diagonal orbit of an $n$-tuple of commuting nilpotent elements of the complex 
group.

Moreover, consider the tangent bundle and suppose $\rho_+$ corresponds to the 
standard representation of $\surmL(n{+}1)$ on $E_{|p}\cong \C^{n+1}$. Then the 
corresponding boundary condition reduces the connection $\ca$ to the 
Levi-Civita connection $\nabla^{\mathrm{LC}}$ as $t \to +\infty$. Thus, on the 
tangent bundle an instanton solution interpolates between the Levi-Civita 
connection and the canonical connection, which is consistent with the findings 
of \cite{harland_noelle}.

Finally, for $\rho_+$ trivial,  one 
observes that the equivariance conditions \eqref{eq:def_equivariance} are 
certainly compatible with both Lie algebra homomorphisms $\rho_\pm$, provided the 
$\widehat{I}_i$ are the images under $\rho_-$ of the $\surmL(n)$ generators.
\subsection{Nilpotent pairs and generalised Nahm's equations}
\label{sec:generalised_Nahm}
The set-up we have encountered for the singular boundary conditions has close 
cousins on both sides: mathematics and physics.

\paragraph{Nilpotent pairs}
Ginzburg introduced the ``doubles'' of nilpotent orbits for semi-simple Lie 
algebras $\gfrak$ in \cite{Ginzburg:2000} and has initiated the study of their 
remarkable properties, which have been further investigated 
\cite{Panyushev:2000,Panyushev:2001,Elashvili:2001}.

Roughly, a \emph{nilpotent pair} $e=(e_1,e_2) \in \gfrak \times \gfrak$ 
satisfies (i) $\com{e_1}{e_2}=0$ and (ii) for an $(t_1,t_2)\in \C^* \times 
\C^*$, there exists $g=g(t_1,t_2) \in G$ such that $(t_1 e_1 ,t_2 e_2) =(\Ad_g 
(e_1),\Ad_g (e_2))$. By this definition, such a pair consists of two commutating nilpotent elements
$e_1$ and $e_2$, but the converse does not necessarily hold.

As noted before, \eqref{eq:def_nilpotent_orbit_diag} is indeed 
an $n$-tuple of commuting nilpotent elements $F_j^-$, see also Appendix 
\ref{app:details}. Hence, we can view it as an example of a natural 
generalisation to something like ``nilpotent tuples''. However, a deeper study 
of these is beyond the scope of this work.
\paragraph{Generalised Nahm's equations}
There exist several generalisations of Nahm's equations in the literature. 
The one that matches our case is the generalisation considered in 
\cite{Xie:2013gma,Hashimoto:2014vpa,Hashimoto:2014nwa,Heckman:2016xdl}, which 
contains two copies of Nahm's equations \eqref{eq:Nahm_eq} coupled by the same 
$\alpha$. 

In \cite{Xie:2013gma,Heckman:2016xdl} the construction of 4-dimensional 
$\Ncal{=}1$ theories from 6-dimensional theories with $\Ncal{=}(2,0)$ or 
$(1,0)$ compactified on 
a Riemann surface with punctures has been studied. Inspired from the dominant 
role of Hitchin equations in 6-dimensional $\Ncal{=}(2,0)$ compactifications to 
4-dimensional $\Ncal{=}2$ theories, the author of \cite{Xie:2013gma} proposed 
generalised Hitchin equations for $\Ncal{=}1$ compactifications, from which 
generalised Nahm's equations have been deduced by reduction. Unsurprisingly, all 
these generalised equations appear as reductions from the 6-dimensional HYM 
equations. Moreover, both equations are important for the space of 
supersymmetric vacua as the moduli space of the generalised Hitchin equations 
describes the Coulomb branch, while the generalised Nahm's equations account for 
the Higgs branch.

Similarly, in \cite{Hashimoto:2014vpa,Hashimoto:2014nwa} $1/4$~BPS 
boundary conditions for 4-dimensional $\Ncal{=}4$ super Yang-Mills theory have 
been 
studied, which again result in 4-dimensional $\Ncal{=}1$ theories.
Recalling the seminal work by Gaiotto and Witten 
\cite{Gaiotto:2008sa,Gaiotto:2008ak}, the study of BPS boundary conditions for 
$\Ncal{=}4$ shed light on the importance of nilpotent orbits via Nahm's 
equations. 
In the case of $1/4$~BPS boundary conditions, generalised Nahm's 
equations appeared in the very same fashion, and resulting moduli spaces have 
to be seen in the context of nilpotent pairs.
A first account of the moduli space of generalised Nahm's equations has been 
given in \cite{Hashimoto:2014nwa} from the GIT quotient perspective.

These instances of generalised Nahm's (and even generalised Hitchin) equations 
agree with our set-up, because all of them are dimensional reductions of HYM 
instanton equations on higher dimensional spaces. Our resulting Nahm-type 
instanton matrix equations arise from an equivariant reduction, which includes
the dimensional reduction.
Hence, studying the new boundary conditions \eqref{eq:bc_matrics_X} extends the 
partial description of \cite{Hashimoto:2014nwa} and formalises generalised 
Nahm's equations to higher dimensions.
Recall that one of the physical origins of our instanton equations are the BPS 
equations for heterotic flux compactifications. 
Therefore, we emphasize the special role of HYM in 6 dimensions as 
generalised Nahm's equations appear as 
(i) BPS equations in heterotic compactifications to $\Ncal{=}1$ in 4d, 
(ii) BPS equations for compactifications of 6-dimensional  $\Ncal{=}(2,0)$ 
theories on Riemann surfaces, and 
(iii) as $1/4$~BPS boundary conditions for 4-dimensional $\Ncal{=}4$ 
super Yang-Mills theory.

In view of \cite{Heckman:2016xdl}, in which solutions to generalised Nahm's 
equations were studied either by reduction to regular Nahm's equations 
or by products of independent $\surm(2)$ subalgebras, we have considered the 
scenario in which 
the tuple of commuting nilpotent elements stems from the orthogonal complement 
$\mfrak$ in the decomposition \eqref{eq:decomp_algebra}. This provides 
another viable option for finding commuting nilpotent elements.
%
%
\section{Instantons on hyper-Kähler cones}
\label{sec:instanton_hk-cone}
In this section we investigate the hyper-Kähler instanton equations for the equivariant ansatz described in Section~\ref{sec:Ansatz_hk}. 

\subsection{Explicit form of hyper-Kähler instanton equations}
\label{sec:explicit_inst_eq}
We now derive the explicit instanton equations on the hyper-Kähler cone by 
evaluating the triplet of 
Hermitian Yang-Mills equations. Denoting, for a fixed $J=J_{\a}$, the holomorphic forms as $\theta^{\a} \eqqcolon e^{a_1}-\im e^{a_2}$ and 
$\theta^{\b} \eqqcolon e^{b_1}-\im e^{b_2}$, the holomorphicity condition $\cf_{\a\b}=0$ reads in terms of real indices 
\begin{align}
\cf_{a_1 b_1} = \cf_{a_2 b_2} \quad \text{and} \quad \cf_{a_1 b_2} \= - 
\cf_{a_2 b_1}.
\end{align}
Using the Kähler forms \eqref{eq:kaehler_forms} on the metric cone, 
we obtain from the first one the conditions
\begin{subequations}
\label{eq:inst_hym}
\begin{alignat}{2}
\label{eq:inst_hym_1}
\nonumber\cf_{4i, 4j+2} &= \cf_{4i+1,4j+3}, \quad & \quad  \cf_{4i,4j+3} &= - 
\cf_{4i+1,4j+2}, \\
\nonumber \cf_{4i, 4j} &= \cf_{4i+1,4j+1}, \quad &  \quad \cf_{4i,4j+1} &= 
-\cf_{4i+1, 4j},\\
 \cf_{4i+2, 4j+2} &= \cf_{4i+3,4j+3},\quad & \quad \cf_{4i+2,4j+3} &= 
-\cf_{4i+3,4j+2},
\end{alignat}
while the second one yields
\begin{alignat}{2}
\label{eq:inst_hym_2}
\nonumber \cf_{4i,4j+3} &= \cf_{4i+2,4j+1}, \quad  & \quad \cf_{4i,4j+1} &= - 
\cf_{4i+2,4j+3},\\
\nonumber \cf_{4i,4j} &= \cf_{4i+2,4j+2}, \quad  & \quad \cf_{4i,4j+2} &= - 
\cf_{4i+2,4j},\\
\cf_{4i+3,4j+3} &= \cf_{4i+1,4j+1}, \quad & \quad \cf_{4i+3,4j+1} &= 
- 
\cf_{4i+1,4j+3},
\end{alignat}
and the third Kähler form   $\Omega_3$ leads to
\begin{alignat}{2}
\label{eq:inst_hym_3}
\nonumber\cf_{4i,4j+1} &= \cf_{4i+3,4j+2}, \quad &  \quad \cf_{4i,4j+2} &= - 
\cf_{4i+3,4j+1},\\
\nonumber \cf_{4i,4j} &= \cf_{4i+3,4j+3}, \quad & \quad \cf_{4i,4j+3} &= - 
\cf_{4i+3,4j},\\
 \cf_{4i+1,4j+1} &= \cf_{4i+2,4j+2}, \quad & \quad \cf_{4i+1,4j+2}  
&= - 
\cf_{4i+2,4j+1}.
\end{alignat}
\end{subequations}
Note that the holomorphicity conditions with respect to any \emph{two} of them already imply the third set of conditions. This is not surprising because 
the characteristic vector fields of two orthogonal Sasaki-structures induce a (unique)
third one and therefore a 3-Sasakian structure, see \cite[Ch.\ 4 Lem.\ 6]{BFGK}.
Moreover, adding the relevant equations for $i=j$ one indeed recovers the three 
stability conditions.

\paragraph{Matrix equations.}
The canonical connection $\widehat{\Gamma}^P$ of \cite{harland_noelle} is by construction an instanton and the 
equivariance condition \eqref{eq:def_equivariance} ensures that there are no 
mixed curvature terms, so that the matrices $X_{\m}$ have to satisfy the 
instanton equations separately. For convenience, we set $X_0=0$ in this 
paragraph. Following \cite[Eq.\ (4.28)]{harland_noelle}, a suitable choice of 
structure constants is given by
\begin{equation}
\label{eq:structure_constants}
 f_{\b\g}^{\a} = - 2 \epsilon_{\b \g}^{\a}, \quad f_{ab}^{\a}= - 2 
\omega_{ab}^{\a}, \quad 
f_{\a b}^{a}= \omega^{\a}_{ab}\; .
\end{equation}

Then one obtains from $i=j=0$ in the instanton conditions \eqref{eq:inst_hym}
the flow equations
for the triplet of  matrices accompanying the contact forms $e^{\a}$,
\begin{equation}
\label{eq:eq:flow_contact}
 \dot{X}_{\a} = - 2 X_{\a} - \tfrac{1}{2} \epsilon_{\a \b \g} \com{X_{\b}}{X_{\g}}
\end{equation}
with $\epsilon_{123}=1$. Setting $i=0$ or $j=0$ yields the flow equations for the other matrices,
\begin{align}
\label{eq:eq:flow_rest}
\nonumber\dot{X}_{4j} &= - X_{4j}+ \com{X_1}{X_{4j+1}}& &= - X_{4j} + \com{X_2}{X_{4j+2}} &
 &= -X_{4j} + \com{X_3}{X_{4j+3}},\\
\nonumber \dot{X}_{4j+1} &=  - X_{4j+1} - \com{X_1}{X_{4j}}&  &= - X_{4j+1} - 
\com{X_2}{X_{4j+3}}& &= -X_{4j+1} + \com{X_3}{X_{4j+2}},\\
\nonumber \dot{X}_{4j+2} &= -X_{4j+2} + \com{X_1}{X_{4j+3}}& &= -X_{4j+2} - 
\com{X_2}{X_{4j}}& &= - X_{4j+2} - \com{X_3}{X_{4j+1}},\\
 \dot{X}_{4j+3} &= -X_{4j+3} - \com{X_1}{X_{4j+2}}& &= - X_{4j+3} + 
\com{X_2}{X_{4j+1}}& &= - X_{4j+3} - \com{X_3}{X_{4j}}.
\end{align}
These flow equation coincide, of course, with the general result given in (3.27) of \cite{ivanova_popov} as we are applying their approach with the same structure
constants. Finally, for $i,j>0$ the instanton equations \eqref{eq:inst_hym} lead to the algebraic relations
\begin{align}
\label{eq:eq_alg1}
\nonumber 4 \delta_{ij} X_1 &= - \com{X_{4i}}{X_{4j+1}} - 
\com{X_{4i+2}}{X_{4j+3}} = - \com{X_{4i}}{X_{4j+1}} - 
\com{X_{4j+2}}{X_{4i+3}},\\
\nonumber 4 \delta_{ij} X_2 &= - \com{X_{4i}}{X_{4j+2}} + 
\com{X_{4i+1}}{X_{4j+3}} = - \com{X_{4i}}{X_{4j+2}} + 
\com{X_{4j+1}}{X_{4i+3}},\\
 4 \delta_{ij} X_3 &= - \com{X_{4i}}{X_{4j+3}} - 
\com{X_{4i+1}}{X_{4j+2}} = - \com{X_{4i}}{X_{4j+3}} - 
\com{X_{4j+1}}{X_{4i+2}},
\end{align}
and
\begin{align} 
\label{eq:eq_alg2}
\nonumber \com{X_{4i}}{X_{4j+1}} &= \com{X_{4j}}{X_{4i+1}},\quad 
\quad \com{X_{4i+2}}{X_{4j+3}} = \com{X_{4j+2}}{X_{4i+3}},\\
\nonumber \com{X_{4i}}{X_{4j+2}} &= \com{X_{4j}}{X_{4i+2}}, 
\quad \quad \com{X_{4i+1}}{X_{4j+3}} = \com{X_{4j+1}}{X_{4i+3}},\\
\com{X_{4i}}{X_{4j+3}} &= \com{X_{4j}}{X_{4i+3}}, \quad \quad 
\com{X_{4i+1}}{X_{4j+2}} = \com{X_{4j+1}}{X_{4i+2}},
\end{align}
as well as
\begin{align}
\label{eq:eq_alg3}
 \com{X_{4i}}{X_{4j}}&= \com{X_{4i+1}}{X_{4j+1}} = 
\com{X_{4i+2}}{X_{4j+2}} = \com{X_{4i+3}}{X_{4j+3}}.
\end{align}
While for $m=0$ the system reduces to the well-known equations on the $\mathrm{SU}(2)$-triplet of contact forms \eqref{eq:eq:flow_contact}, for any positive 
$m$ the system gets significantly more complicated due to the occurrence of the 
non-trivial algebraic relations \eqref{eq:eq_alg1}. In particular, the three 
matrices $X_{\a}$ can be expressed as commutators of the other matrices, so that the flow equations for $X_{a}$
are actually \emph{cubic} in the endomorphisms. Moreover, by virtue of these algebraic conditions, the flow equations for $X_{\a}$, $\a=1,2,3$, follow from
the flow equations \eqref{eq:eq:flow_rest} of the other matrices.

We will comment on this algebraic behaviour, different from that of instanton 
equations for a single
Sasaki-Einstein structure as in \cite{Sperling:2015sra}, in more detail in the following section.
\subsection{Single HYM moduli space}
The discussion of Section \ref{sec:equivalence_Sp_HYM} has shown that 
$\sprm(m)$-instantons on hyper-Kähler cones over 3-Sasakian manifolds
can be described as the intersection of the holomorphicity conditions with respect to the $\C P^1$ family of Sasaki-Einstein structures. 
As shown in the previous section, it is even
sufficient to consider only the intersection of the HYM equations with respect to \emph{two} orthogonal Sasaki-Einstein structures. 

Therefore, we give a description of such a HYM space here, commenting also on 
the differences compared to \cite{Sperling:2015sra}
due to the different algebraic conditions. Without loss of generality, let us 
specialise to $\Omega=\Omega_3$ in the following.

Conceptually, we need to adjust the setting compared to 
Section \ref{sec:instanton_cy-cone}:
the starting point $\Gamma^P$ is the canonical connection for the 
$\sprm(m)$-structure on the base, which is an $\surm(2m{+}1)$-instanton due to 
$\sprm(m) \subset \surm(2m) \subset \surm(2m{+}1)$. Nevertheless, the more
``natural'' starting point would have been the canonical instanton in the sense of \cite{harland_noelle} associated 
to the Sasaki-Einstein $\surm(2m{+}1)$-structure, as it has been used for the Calabi-Yau cones in \cite{Sperling:2015sra}.

\paragraph{Holomorphicity condition.} The holomorphicity conditions of the HYM 
equations for $\Omega_3$ yield the differential equations
\begin{align}
\label{eq:hym3_flow}
\nonumber \dot{X_1} &= -\com{X_2}{X_3} - 2 X_1 &  \dot{X_2} &= -\com{X_3}{X_1} 
- 2 X_2,\\
\nonumber          \dot{X}_{4i+1} &= \com{X_3}{X_{4i+2}} - X_{4i+1} & \dot{X}_{4i+2} &= 
-\com{X_3}{X_{4i+1}} - X_{4i+2},\\
 \dot{X}_{4i} &= \com{X_3}{X_{4i+3}} - X_{4i} & \dot{X}_{4i+3} &= - 
\com{X_3}{X_{4i}} - X_{4j+3},
\end{align}
together with the algebraic relations (for $i,j>0$)
\begin{align}
\label{eq:hym3_alg1}
\nonumber
 &\com{X_1}{X_{4i+1}} = \com{X_2}{X_{4i+2}} & &\com{X_1}{X_{4i+2}} = 
-\com{X_2}{X_{4i+1}},\\
&\com{X_1}{X_{4i}}=\com{X_2}{X_{4i+3}}, & &\com{X_1}{X_{4i+3}}=- 
\com{X_2}{X_{4i}}.
\end{align}
and
\begin{align}
\label{eq:hym3_alg2}
\nonumber
 4 \delta_{ij} X_1 &= \com{X_{4i+3}}{X_{4j+2}} - \com{X_{4i}}{X_{4j+1}} & 4 
\delta_{ij}X_2 &= - \com{X_{4i+3}}{X_{4j+1}} - \com{X_{4i}}{X_{4j+2}},\\
\nonumber
 0& = \com{X_{4i}}{X_{4j}}-\com{X_{4i+3}}{X_{4j+3}} & 0&=\com{X_{4i}}{X_{4j+3}} 
-\com{X_{4j}}{X_{4i+3}},\\
0& = \com{X_{4i+1}}{X_{4j+1}} -\com{X_{4i+2}}{X_{4j+2}} & 0 
&=\com{X_{4i+1}}{X_{4j+2}}-\com{X_{4j+1}}{X_{4i+2}}.
\end{align}
Again, the algebraic conditions combined with the differential equations of 
$X_a$ for $a=4, \ldots, 4m+3$ imply the differential equations
for the endomorphisms $X_1$ and $X_2$.

\paragraph{Stability-like condition.} Evaluating $\Omega_3 \lrcorner \cf=0$ for 
the given form $\Omega_3$ and the structure constants 
\eqref{eq:structure_constants}
leads to
\begin{align}
\label{eq:stab_cond}
\begin{aligned}
- \dot{X}_3 &= \com{X_1}{X_2} +2X_3+ \sum_{i=1}^m 
\Big( \com{X_{4i+1}}{X_{4i+2}}+2X_3 +\com{X_{4i}}{X_{4i+3}}+2X_3 \Big)  \\
 &= \com{X_1}{X_2} + 2(2m+1) X_3 + \sum_{i=1}^m 
\Big(\com{X_{4i+1}}{X_{4i+2}} +\com{X_{4i}}{X_{4i+3}}\Big) \, .
\end{aligned}
\end{align}
As in the previous discussions of the Nahm-type equations,  the flow equations \eqref{eq:hym3_flow} will be referred to as 
\emph{complex equations}, while the stability-like condition 
is the \emph{real equation}.
\paragraph{Gradient flow.}
The differential equations \eqref{eq:hym3_flow} and \eqref{eq:stab_cond} among the 
instanton matrix equations  can be cast as gradient flow equations $\frac{\diff 
}{\diff t} X = - \nabla \Phi(X)$ similarly to 
Nahm's equations \cite{Kronheimer:1990ay}. To see this, let $\Phi : 
\urmL(p)^{\times (4m{+}3)} \to \R$ be the function defined as
\begin{align}
\label{eq:gradient_flow}
\begin{aligned}
 \Phi(X_\alpha,X_a)\coloneqq &\tr(X_1 X_1) + \tr(X_2 X_2) + (2m+1) \tr(X_3 X_3) 
+ \frac{1}{2} \sum_{a=4}^{4m+3} \tr(X_a X_a) \\
&+\tr\left(X_3 \com{X_1}{X_2}\right)
+\tr\left(X_3 \sum_{j=1}^{m}\left(  \com{X_{4j+1}}{X_{4j+2}} + 
\com{X_{4j}}{X_{4j+3}} \right) \right)\; .
\end{aligned} 
\end{align}
The algebraic equations \eqref{eq:hym3_alg1}, \eqref{eq:hym3_alg2} are not part 
of this system, but they are invariant under the gradient flow. Thus, if they 
are satisfied at any point $t_0$ then they hold throughout the evolution.
\paragraph{Rewriting of the matrix equations.}
One can eliminate the linear terms in the instanton matrix equations \eqref{eq:hym3_flow} and \eqref{eq:stab_cond} by a 
suitable rescaling as follows:
\begin{align}
\label{eq:rescaling_fields}
 X_{\g}=\e^{-2(2m+1)\tau} \Xcal_{\g}\ \ (\g=0,3),  
\quad X_{\b}= \e^{-2\tau}\Xcal_{\b}\ \ (\b=1,2), \quad
  \text{and} \quad
X_{a}= \e^{-\tau}\Xcal_a
\end{align}
for $a=4, \ldots 4m+3$, which is accompanied by a rescaled cone coordinate
\begin{align}
s \coloneqq -\frac{1}{2(2m+1)}\e^{-2(2m+1)\tau}.
\end{align}
In addition, we combine the matrices into the complex fields (defined w.r.t. $J_3$),
\begin{align}
\label{eq:def_cplx_matrices}
\begin{aligned}
\Pcal_i &\coloneqq \tfrac{1}{2}(\Xcal_{4i+1}+\im \Xcal_{4i+2}),  &\qquad 
\Qcal_i &\coloneqq \tfrac{1}{2}(\Xcal_{4i}  +\im \Xcal_{4i+3}), \\
\Ycal &\coloneqq \tfrac{1}{2}(\Xcal_1 + \im \Xcal_2), & \quad 
\Zcal &\coloneqq \tfrac{1}{2}(\Xcal_0+  \im \Xcal_3).
\end{aligned}
\end{align}
The complex equations then read as follows:
the purely algebraic relations \eqref{eq:hym3_alg1} and \eqref{eq:hym3_alg2}
\begin{subequations}
\label{eq:cplx_eq}
\begin{align}
\label{eq:cplx_eq_alg}
\com{\Pcal_i}{\Pcal_j} = 0 \= \com{\Qcal_i}{\Qcal_j}, 
\quad \quad 
\com{\Pcal_i}{\Ycal} \= 0 \= \com{\Qcal_i}{\Ycal}, \quad 
\quad
\com{\Pcal_i}{\Qcal_j} \= 2\delta_{ij} \Ycal
\end{align} 
for $i,j=1, \ldots m$, which are the commutation relations of a (complexified) 
Heisenberg algebra.
Moreover, the differential equations simplify to
\begin{align}
\label{eq:cplx_eq_diff}
\frac{\diff}{\diff s} \Pcal_i = 2 \com{\Pcal_i}{\Zcal}, 
\quad \quad \frac{\diff}{\diff s} \Qcal_i \= 2 
\com{\Qcal_i}{\Zcal}, 
\quad \quad \frac{\diff}{\diff s} \Ycal \= 2 
\com{\Ycal}{\Zcal}.
\end{align}
\end{subequations}
The real equation becomes
\begin{subequations}
\label{eq:real_eq}
\begin{align}
\frac{\diff}{\diff s} (\Zcal + \Zcal^\dagger)  
+ 2 \com{\Zcal}{\Zcal^\dagger} 
+ \l_1(s) \com{\Ycal}{\Ycal^{\+}}                        
+\l_2(s)\sum_{i=1}^{m} \left(\com{\Qcal_i}{\Qcal_i^{\+}}			
 +\com{\Pcal_i}{\Pcal_i^{\+}} \right) =0
\end{align}
with the functions
\begin{align}
\l_1(s) \coloneqq \lb -\frac{1}{2(2m+1)s}\rb^{\tfrac{2(2m+1)-2}{2m+1}} \quad 
\text{and} \quad
\l_2(s) \coloneqq \lb -\frac{1}{2(2m+1)s}\rb^{\tfrac{2(2m+1)-1}{2m+1}}.
\end{align}
\end{subequations}
Note that the definition of $s$ coincides with that in \cite{Sperling:2015sra} 
(recalling $n=2m+1$ in that notation), but that the functions $\l$ differ 
because in the hyper-Kähler case only the rescaling of $X_3$ depends on $m$ 
while the other factors of $\tau$ in \eqref{eq:rescaling_fields} are the same  for all dimensions $m$.

Similarly to the remarks in Section~\ref{sec:generalised_Nahm}, one may think of the above equations
as $m$ copies of the same system, which all have the same matrix $\mathcal{Y}$. This, of course, just reflects the geometry
of the hyper-K\"ahler manifold as consisting of quaternionic ``blocks'' on which the defining structures act. 

\paragraph{Preliminaries.}
Just to be explicit, we recall the gauge transformations for complex matrix 
equations. The real and the complex group of gauge transformations (with respect 
to $J_3$) are 
respectively
\begin{align}
 \Ggauge_{\mathrm{HYM}}= \{g : \R^- \to \surm(p)\} \; , \qquad
  \Ggauge_{\mathrm{HYM}}^\C= \{g : \R^- \to \slrm(p,\C)\} \; .
\end{align}
The transformation rules of the complex linear combinations 
\eqref{eq:def_cplx_matrices} are
\begin{alignat}{2}
 \Pcal_j &\mapsto \Pcal_j^g \coloneqq \Ad(g) \Pcal_j \; , & \qquad
 \Qcal_j &\mapsto \Qcal_j^g \coloneqq \Ad(g) \Qcal_j  \quad \text{for } 
j=1,\ldots,m \; ,\\
 \Ycal &\mapsto \Ycal^g \coloneqq \Ad(g) \Ycal \; , & \qquad 
 \Zcal &\mapsto \Zcal^g \coloneqq \Ad(g) \Zcal - \tfrac{1}{2} \left( 
\frac{\diff}{\diff s} g \right) g^{-1} \; .
\end{alignat}
We emphasize that the complex equations are invariant under the complex gauge 
transformations, while the real equation is only invariant under the real gauge 
transformations.
\paragraph{Formulation of boundary conditions.}
As discussed in Section \ref{sec:instanton_cy-cone}, the generic model solution 
is of the form
\begin{subequations}
\begin{align}
X_{a} &= \e^{-\tau}\Tcal_a + \Scal_a, \quad a=4, \ldots 4m+3 \, , \\
X_{\b}&= \e^{-2\tau}\Tcal_{\b} + \Scal_{\b}, \quad \b=1,2 \, , \\
X_{3} &=\e^{-2(2m+1)\tau} \Tcal_3 + \Scal_3, \; \text{and} \; X_{0}=0  \, ,
\end{align}
\end{subequations}
where the $\Tcal$ are a solution to

\begin{subequations}
\label{eq:trivial_solutions}
\begin{align}
 \com{\Pcal_i}{\Pcal_j} &= 0 \= \com{\Qcal_i}{\Qcal_j} \; , \\
\com{\Pcal_i}{\Ycal} &= 0 \= \com{\Qcal_i}{\Ycal}, \qquad
\com{\Pcal_i}{\Qcal_j} \= 2\delta_{ij} \Ycal \; ,\\
  \com{\Pcal_i}{\Zcal} &= 
0 =  \com{\Qcal_i}{\Zcal}
 =   \com{\Ycal}{\Zcal} \; ,
\end{align}
\end{subequations}
wherein the complex linear combinations \eqref{eq:def_cplx_matrices} are formed 
out of the $\Tcal$.

The obvious observation is $\Zcal$ commutes with every other matrix. Next,
$\Pcal_i$ and $\Qcal_j$ commute with each other and among themselves; thus, 
resembling the complexified algebra of $\R^{2m}$. However, $\Ycal$ introduces a 
central extension, which renders the algebra spanned by $\Pcal_i,\Qcal_j, 
\Ycal$ into a complexified Heisenberg algebra $H_m^\C$.
The solution space to \eqref{eq:trivial_solutions} is not empty, 
because the choice $\Ycal\equiv0$ allows all other generators to be 
chosen from a Cartan subalgebra of $\glrmL(p,\C)$.

Next, the matrices $\Scal$ commute with all $\Tcal$ and are critical points of 
$\Phi$, see \eqref{eq:gradient_flow}, subject to the additional constraints 
\eqref{eq:hym3_alg1}, \eqref{eq:hym3_alg2}. In detail, $\Scal$ need to satisfy 
the following commutation relations:
\begin{subequations}
\label{eq:constant_offset}
\begin{align}
\com{\Scal_2}{\Scal_3} &= - 2 \Scal_1 \, , &  
\com{\Scal_3}{\Scal_1}  &= - 2 \Scal_2 \, , \\
\com{\Scal_3}{\Scal_{4i+2}}&=   \Scal_{4i+1} \, , & 
\com{\Scal_3}{\Scal_{4i+1}} &=  - \Scal_{4i+2} \, , \\
\com{\Scal_3}{\Scal_{4i+3}} &=   \Scal_{4i} \, , & 
\com{\Scal_3}{\Scal_{4i}}  &= - \Scal_{4j+3} \, , \\
\com{\Scal_1}{\Scal_{4i+1}} &= \com{\Scal_2}{\Scal_{4i+2}} \, , & 
\com{\Scal_1}{\Scal_{4i+2}} &= 
-\com{\Scal_2}{\Scal_{4i+1}} \, , \\
\com{\Scal_1}{\Scal_{4i}} &=\com{\Scal_2}{\Scal_{4i+3}} \, , & 
\com{\Scal_1}{\Scal_{4i+3}}&=- \com{\Scal_2}{\Scal_{4i}} \, ,
\end{align}
and
\begin{align}
4 \delta_{ij} \Scal_1 &= \com{\Scal_{4i+3}}{\Scal_{4j+2}} - 
\com{\Scal_{4i}}{\Scal_{4j+1}}  \, , & 4 
\delta_{ij}\Scal_2 &= - \com{\Scal_{4i+3}}{\Scal_{4j+1}} - 
\com{\Scal_{4i}}{\Scal_{4j+2}} \, , \\
0& = \com{\Scal_{4i}}{\Scal_{4j}}-\com{\Scal_{4i+3}}{\Scal_{4j+3}}  \, , & 
0&=\com{\Scal_{4i}}{\Scal_{4j+3}} 
-\com{\Scal_{4j}}{\Scal_{4i+3}} \, ,\\
0& = \com{\Scal_{4i+1}}{\Scal_{4j+1}} -\com{\Scal_{4i+2}}{\Scal_{4j+2}} \, , & 
0 
&=\com{\Scal_{4i+1}}{\Scal_{4j+2}}-\com{\Scal_{4j+1}}{\Scal_{4i+2}} \, ,
\end{align}
as well as
\begin{align}
0 = 
\com{\Scal_1}{\Scal_2} + 2(2m+1) \Scal_3 + \sum_{i=1}^m 
(\com{\Scal_{4i+1}}{\Scal_{4i+2}} +\com{\Scal_{4i}}{\Scal_{4i+3}})\, .
\end{align}
\end{subequations}
However, in contrast to the case of Nahm's equations \cite{Kronheimer:1990ay} 
the 
critical points of $\Phi$ do not necessarily give rise to a Lie algebra 
homomorphism, unless one considers the trivial case $m=0$, which reduces to the set-up of the original Nahm's equations, of course.

Comparing to \cite{Sperling:2015sra,Sperling:2016dqk}, one 
could impose that the $\Tcal$ are a regular\footnote{the intersection of the 
centralisers of the $T_\mu$ consists  only of a Cartan subalgebra} tuple, 
but one would necessarily have to set $\Ycal=0$. For boundary conditions with 
non-vanishing $\Ycal$, regularity cannot be required. Consequently, one cannot 
dismiss the possibility of 
having non-trivial $\Scal$ in the boundary conditions. Therefore, the generic 
solution to the matrix equations is determined by two tuples $\Tcal$ and $\Scal$ 
of matrices in the boundary conditions.

The situation is in analogy to Nahm's equations considered in 
\cite{kronheimer1990hyper,Kronheimer:1990ay,Biquard:1996,kovalev}. While 
Kronheimer 
studied the ``extreme'' cases $\Tcal=0$ or $\Scal=0$, Biquard and Kovalev considered 
generic boundary conditions. In all cases, the idea has been to assign suitable 
boundary conditions to the Nahm equations, for which the moduli space is known 
to be hyper-Kähler due to Hitchin, and conclude that general coadjoint orbits 
of complexified Lie groups are hyper-Kähler.
Here, we aim for less: learn as much as possible about the HYM 
matrix instanton equations by generalising this analysis, 
because the matrix equations exhibit a Nahm-type structure. 

In the light of Section \ref{sec:instanton_cy-cone}, we expect that the 
solutions to the complex equations are classified by a ``diagonal'' 
generalisation of a general coadjoint orbit. This is similar to 
$\Ocal_{\diag}$ of \eqref{eq:def_adj_orbit_diag} and $\Ncal_{\diag}$ of 
\eqref{eq:def_nilpotent_orbit_diag}. The analysis is expected to follow the 
arguments of \cite{Biquard:1996}. In other words, one first considers the 
complex equations for suitable boundary conditions. Most arguments from Section 
\ref{sec:instanton_cy-cone} and \cite{Sperling:2015sra} still hold, only the 
local solution has to be adapted. Thus, we expect that the conjugacy classes of 
the ``complex  trajectories'' can be identified with a suitable orbit. 
Secondly, the analysis of the real equations remains the same.
For our intents and purposes, it therefore suffices to note that the moduli 
space of the HYM matrix equations has a Kähler structure and is mapped into some 
finite-dimensional orbit space.

However, despite the formal similarities in the description of a \emph{single} HYM moduli space contained 
in the description of the $\mathrm{Sp}(m)$ instantons, one should keep in mind that the intersection
 \eqref{eq:intersection_mod_space} is very restrictive: Since the triplet $X_{\a}$ has to transform the same 
and due to the way they couple to each other in the matrix equations, there is no (obvious) gauge transformation 
for all Kähler structures simultaneously in the case $m \geq 1$. Therefore, one 
may need completely new tools, taking into account 
the $\mathrm{SU}(2)$-symmetry of the fibre and the quaternionic structures of 
the other matrices explicitly, to describe the generic
properties of the moduli space.

%
%
%
\subsection{Space of equivariant connections}
\label{sec:equivariance}
Due to the choice of structure constants \eqref{eq:structure_constants} and 
their intimate relationship to the hyper-Kähler structure forms 
\eqref{eq:kaehler_forms}, it is not surprising that  one can express the 
equivariance conditions as holomorphic equations in the matrix valued-functions 
for a given complex structure $\bJ_\alpha$. The relation to the complex 
structure on the metric cone over $\mfd{4m+3}$ is established via 
\eqref{eq:def_cplx_str_connection}.

The exact description of the equivariant connections depends on the concrete 
3-Sasakian manifold taken into account, but we review the example of the 
squashed 
seven-sphere, as studied in \cite{s7}. Its equivariance condition requires
$ \com{\widehat{I}_j}{X_{\m}} \= f_{j\m}^{\n} X_{\n}$,
where $j=8,9,10$  labels the generators of the $\sprm(1)$ subgroup. The 
relevant non-vanishing structure constants are
\footnote{By mapping the indices 
$(e^{1},e^{2},e^{3},e^{4},e^{5},e^{6},e^{7}) 
\mapsto  (e^4, e^5, e^6, e^7, -e^2, -e^3, e^1)$ we relate our notation with 
that of \cite{s7}.}
\begin{equation}
\begin{aligned}
-f^4_{85}&= f^4_{96} =f^4_{10,7} \=1\, , \quad & \quad 
f^5_{84} &= f^5_{97} = - f^5_{10,6} =1 \, ,\\
 f^6_{87}&=-f^6_{94} = f^6_{10,5} =1 \, , \quad & \quad -f^7_{86} &= 
-f^7_{95}=-f^7_{10,4}=1 \, .
\end{aligned} 
\end{equation}
and the equivariance conditions therefore read
\begin{equation}
\begin{aligned}
\com{\widehat{I}_8}{X_4}&= X_5 \, , & \com{\widehat{I}_8}{X_5}&= -X_4 \, , &  
\com{\widehat{I}_8}{X_6} &= -X_7 \, , &  \com{\widehat{I}_8}{X_7} &= X_6 \, ,\\
\com{\widehat{I}_9}{X_4}&= -X_6 \, , & \com{\widehat{I}_9}{X_5}&= -X_7\, , &  
\com{\widehat{I}_9}{X_6} &= X_4 \, , &  \com{\widehat{I}_9}{X_7} &= X_5 \, , \\
\com{\widehat{I}_{10}}{X_4}&= -X_7 \, , & \com{\widehat{I}_{10}}{X_5}&= X_6 \,, 
& \com{\widehat{I}_{10}}{X_6} &= -X_5 \, , &  \com{\widehat{I}_{10}}{X_7} &= 
X_4 \, ,
\end{aligned} 
\end{equation}
and
\begin{align}
 \com{\widehat{I}_j}{X_{\a}} &=0 \quad \quad \forall j=8,9,10, \ \a=1,2,3 \,.
\end{align}
The complex structures act according to $J_{\a}$ in \eqref{eq:action_j} on the tangent vectors 
$\delta X_{\m}$. 
Imposing the equivariance condition is compatible with the hyper-Kähler 
structure given by $J_{\a}$ because the equations are invariant. Consider for 
instance 
\begin{align}
\com{\widehat{I}_8}{X_4} = X_5 \qquad \Rightarrow \qquad \com{\widehat{I}_8}{\delta X_4} = 
\delta X_5.
\end{align}
Applying $J_1$ gives us
\begin{align}
\com{\widehat{I}_8}{J_1 (\delta X_4)} \= \com{\widehat{I}_8}{-\delta X_5}\= - (-\delta X_4) \= J_1 
(\delta X_5) \= J_1 \com{\widehat{I}_8}{(\delta X_4)}.
\end{align}
Similarly, we have
\begin{align}
\nonumber
\com{\widehat{I}_8}{J_2 (\delta X_4)} &= \com{\widehat{I}_8}{-\delta X_6} \= - (-\delta X_7) \= J_2 
(\delta X_5)\= J_2 \com{\widehat{I}_8}{(\delta X_4)}\\
\com{\widehat{I}_8}{J_3 (\delta X_4)} &= \com{\widehat{I}_8}{-\delta X_7} \= - \delta X_6 \= J_3 
(\delta X_5)\= J_3 \com{\widehat{I}_8}{(\delta X_4)}. 
\end{align}
Thus, the space of equivariant connections $\Aa^{\mathrm{equiv}}$ is a
tri-holomorphic subspace of $\Aa^{\mathrm{holo}}$. The remaining question is 
whether or not the metric or equivalently the symplectic structure is 
non-degenerate on the vanishing locus of \eqref{eq:def_equivariance}. It seems 
difficult to obtain an exact statement for the generic case.
%
%
\section{Summary and conclusions}
\label{sec:conclusion}
In the course of this article we considered higher-dimensional instantons on 
Calabi-Yau cones and hyper-Kähler cones over arbitrary Sasaki-Einstein and 
3-Sasakian manifolds $\mfd{k}$, respectively.
It is known that the instanton moduli space 
over a (hyper-)Kähler manifold is (hyper-)Kähler, resulting from an 
infinite-dimensional (hyper-)Kähler quotient. It seems naturally that the 
subset of invariant connections inherits this property, but the overall 
situation remains unknown.

In the ansatz \eqref{eq:ansatz_gauge_conn} presented, we restricted the 
connections to those obtained by extension of the (lifted) canonical connection 
$\Gamma^P$ on $T\mfd{k}$ by $t$-dependent endomorphisms-valued 1-forms 
$X_\mu(t) \otimes e^\mu$ which satisfy an equivariance condition 
\eqref{eq:def_equivariance}.
For this ansatz we specified the geometric structures on the space of 
connections.

In Section \ref{sec:instanton_cy-cone} we have significantly extended the 
study of the Nahm-type instanton matrix equations on Calabi-Yau cones. On the 
one hand, we extended the discussion of the regular boundary conditions started 
earlier in \cite{Sperling:2015sra} by providing details of the relevant diagonal 
coadjoint orbit \eqref{eq:def_adj_orbit_diag}. On the other hand, we have 
conducted the full treatment of boundary conditions given by Lie algebra 
homomorphisms. 
As in the study of $4$-dimensional instantons, these boundary 
conditions seem to be the most physical, as they relate to known instantons, for 
instance, on the tangent bundle. 
Similar to Kronheimer's case, the moduli space is related to a 
``diagonal'' nilpotent orbit \eqref{eq:def_nilpotent_orbit_diag} or, 
equivalently, an orbit of an $n$-tuple of commutating nilpotent elements.

These generalised Nahm's equations appear in the construction of 4-dimensional 
$\Ncal{=}1$ theories from heterotic string theory, 6-dimensional 
gauge theories, or 4-dimensional theories with higher amount of supersymmetry. 
We complemented the study of their moduli space, started in 
\cite{Hashimoto:2014vpa,Hashimoto:2014nwa}, and generalised the system to the 
reduction obtained from $2n$-dimensional HYM-equations, $n\geq3$. In addition, 
the ansatz taken represents a complementary ansatz compared to 
\cite{Heckman:2016xdl}.
The treatment of generalised Nahm's equations for $n\geq3$ suggests a 
close relationship of their moduli space with orbits of $n$-tuples of commuting 
nilpotent elements, which we proposed as natural extension of nilpotent pairs 
introduced in \cite{Ginzburg:2000}. However, since the classification of 
commuting nilpotent pairs is, up to our knowledge, still an open problem, we 
refrain from any speculation about nilpotent $n$-tuples.

Using the equivalence between $\sprm(m)$-instantons and a $\C P^1$-family of 
HYM instantons, we described explicitly the system of HYM instanton matrix 
equations of a single $\surm(2m)$-structure in 
Section \ref{sec:instanton_hk-cone}. 
Due to the different starting point
$\Gamma^P$ on the 3-Sasakian base, the equations behaved differently compared 
to the Sasaki-Einstein canonical connection of the Calabi-Yau cone of 
Section~\ref{sec:instanton_cy-cone}.
 Nonetheless, the overall picture remains: the 
Nahm-like equations are expected to have a Kähler structure on the moduli space 
and the precise treatment of boundary conditions only changes the orbit into 
which the space is embedded to. 

However, the structure of the entire intersection 
of the single HYM moduli spaces is not yet fully understood. The complications 
can be traced back to the different bundle 
structure of 3-Sasakian manifolds as $\mathrm{SU}(2)$ (or $\mathrm{SO}(3)$)-bundle, while in the usual Sasaki-Einstein case one had 
a $\mathrm{U}(1)$-bundle over the underlying space. The latter allowed for a 
direct generalization of the Nahm-type equations on Calabi-Yau cones
\cite{Sperling:2015sra}, while the complete discussion of the hyper-Kähler case 
may require new
approaches and is left for future work. By virtue of the absolutely regular formulation of hyper-K\"ahler instantons, one expect 
a generic description for all $m$, once the case of $m=1$ is understood, similarly to the generic results obtained in \cite{Sperling:2015sra}. 
%
%
%
\paragraph{Acknowledgements.}
We are grateful to Fabio Apruzzi, Felix Lubbe, Alexander D. Popov, and Markus 
Röser for valuable discussions and comments. This work was done within the 
framework of the DFG project LE 838/13. JG is supported by the DFG research 
training group GRK1463 ``Analysis, Geometry, and String Theory''. 
MS is supported by Austrian Science Fund (FWF) grant P28590. 
\appendix
\section{Details on non-regular boundary conditions}
\label{app:details}
We provide the details for adaptation of Kronheimer's treatment of the Nahm's 
equations in \cite{Kronheimer:1990ay}; in particular, focusing on 
\cite[Lem.\ 10 \& Lem.\ 11]{Kronheimer:1990ay}.
\paragraph{Preliminaries.}
For a  semi-simple Lie algebra of rank $k$ with simple roots $\alpha_i$, 
$i=1,\ldots,r$,
we recall the Chevalley basis 
\begin{subequations}
\begin{align}
 \com{H_{\alpha_i}}{H_{\alpha_j}} &=0 \; ,\\
 \com{H_{\alpha_i}}{E_{\alpha_j}} &=A_{ji} E_{\alpha_j}  \; ,\\
 \com{ E_{-\alpha_i}}{E_{\alpha_i}} &= H_{\alpha_i}  \; ,\\
 (\ad_{E_{\pm \alpha_i}})^{1-A_{ji}} E_{\pm \alpha_j} &= 0 \; .
\end{align}
\end{subequations}
Here, $A_{ji}$ denotes the Cartan matrix elements.
The last line, the \emph{Serre relations}, imply that 
$\com{E_{\alpha_i}}{E_{\alpha_j}}$ is non-vanishing 
only if $\alpha_i + \alpha_j$ is a root.
Then the split $\surmL(n+1) = \surmL(n)\oplus\mfrak$ can be expressed in terms 
of 
a Cartan subalgebra (CSA) and roots as shown in Table \ref{tab:roots_An}. 
\begin{table}[h]
\centering
\begin{tabular}{c|c|c|c}
\toprule
  & $\surmL(n+1)$ &  $\surmL(n)$ & $\mfrak$ \\ \midrule
  CSA & $H_{\alpha_i}$, $i=1, \ldots, n$ & 
$H_{\alpha_i}$, $i=1, \ldots, n-1$ &
$H_{\alpha_n}$ \\
simple roots & $E_{e_i - e_{i+1}}$, $i=1, \ldots, n$ & 
$E_{e_i - e_{i+1}}$,  $i=1, \ldots, n-1$ &
$E_{e_n - e_{n+1}}$ \\
positive roots & $E_{e_i - e_{j}}$, $1\leq i < j \leq n+1$ & 
$E_{e_i - e_{j+1}}$, $1\leq i < j \leq n$ &
$ E_{e_i - e_{n+1}}$, $i=1,\ldots,n$ \\
\bottomrule
\end{tabular}
\caption{The roots are given in terms of the ONB $e_i$ on $\R^{n+1}$. See for 
instance \cite{Fuchs:1997jv}.}
\label{tab:roots_An}
\end{table}
Recall that $E_{e_i - e_{i+1}}$ are the ``creation operators'' for the 
$\surmL(2)$ subalgebra spanned by $\{H_{\alpha_i}, E_{\pm(e_i - e_{i+1})}\}$. 
The important question for later is whether the ``creation operators'' $ E_{e_i 
- e_{n+1}}$, $i=1,\ldots,n$ on $\mfrak$ commute with each other.
From the Serre relations we observe 
\begin{align}
 \com{ E_{e_i - e_{n+1}}}{  E_{e_j - e_{n+1}}} &=0 
\end{align}
because $(e_i - e_{n+1}) + (e_j - e_{n+1}) = e_i +e_j -2 e_{n+1}$ is not 
a root.

Inspired from the explicit calculations for $\surm(3)\slash \surm(2)$ in 
\cite{Lechtenfeld:2015ona} and $\surm(4) \slash \surm(3)$ in \cite{s7}, we can 
replace $H_{\alpha_n} \in \mfrak$ by a new element $\tilde{H}$ such that
\begin{align}
 \ad_{\tilde{H}}(E_{e_i - e_{n+1}}) = (n+1) E_{e_i - e_{n+1}} \qquad \forall i 
=1,\ldots,n \; .
\end{align}
Additionally, one can rescale $\tilde{H}$ as 
\begin{align}
 \tilde{H} \mapsto H = \frac{1}{n}\tilde{H}
\end{align}
such that 
\begin{align}
 \ad_{H}(E_{e_i - e_{n+1}}) = \frac{n+1}{n} E_{e_i - e_{n+1}} \qquad 
\forall i =1,\ldots,n \; .
\end{align}
This is the same rescaling as employed in the definition of the torsion 
components of the canonical connection $\Gamma^P$ of \cite{harland_noelle}. See 
also \cite{Lechtenfeld:2015ona} for an explicit example in $n=2$.
From now on denote
\begin{align}
 E_j \coloneqq E_{e_j - e_{n+1}} \; , \qquad F_j \coloneqq E_{-(e_j - e_{n+1})} 
\; ,
\label{eq:def_nilpotent_elements}
\end{align}
and note that the $E_j$ are nilpotent.\\

 We now discuss the adaptation of Kronheimer's ``complex trajectories'' and Lemma~10 and 
Lemma~11 from \cite{Kronheimer:1990ay} to generic Calabi-Yau cones:

\paragraph{Adaptation of ``complex trajectory''.}
Let $\rho_{\pm} : \surmL(n{+}1) \to \glrmL(p,\C)$ be two Lie algebra 
homomorphisms and denote the images of the $\surmL(n{+}1)$ generators $H$ and $E_j$ from above as $H^{\pm}$ and $E_j^{\pm}$.
Then a \emph{complex trajectory} is an $(n+1)$-tuple of smooth 
functions $(Y_{n+1},Y_j) : \R \to \glrmL(p,\C)$ such that
\begin{enumerate}[(i)]
\item  the complex equations \eqref{eq:cplx_eq_CY} are satisfied,
 \item for $t\to + \infty$ 
 \begin{subequations}
 \label{eq:def_cplx_trajectory}
 \begin{align}
  2 Y_{n+1} (t) \to H^+ \, , \qquad 
  Y_j(t) \to E_j^+ \, , \quad \forall j=1, \ldots, n \; , \label{eq:bc_+infty}
 \end{align}
 \item and for $t\to - \infty$ 
 \begin{align}
  2 Y_{n+1} (t) \to \Ad_g (H^-) \, , \qquad 
  Y_j(t) \to \Ad_g(E_j^-) \, , \quad  \forall j=1, \ldots, n
  \label{eq:bc_-infty}
 \end{align}
 \end{subequations}
for some $g\in \urm(p)$ in the compact group.
\end{enumerate}

Two complex trajectories $(Y_{n+1},Y_j)$, $(Y'_{n+1},Y'_j)$ are 
\emph{equivalent} if there exists a map $g:\R \to \glrm(p,C)$ with $g\to 1$ as 
$t\to \infty$, such that $(Y'_{n+1},Y'_j) = g (Y_{n+1},Y_j) $.
Similarly to \cite[Lem.\ 9]{Kronheimer:1990ay}, if two complex trajectories are 
equal outside a compact subset of $\R$, then they are equivalent in the above 
sense.
\paragraph{Adaptation of Lemma 10.}
Let $(Y_{n+1},Y_j)$ be a solution of the complex equations \eqref{eq:cplx_eq_CY}
satisfying the boundary conditions \eqref{eq:bc_-infty}. Then there exists a 
gauge transformation $g_-:\R \to \glrm(p,\C)$ 
with $\lim_{t\to -\infty} g_- = \mathrm{const.}$ such that 
$(Y'_{n+1},Y'_j)=g_-(Y_{n+1},Y_j)$ is a constant solution
\begin{align}
 2 Y'_{n+1} = H^- \;, \qquad Y'_j = F_j^- \; .
 \end{align}

To see this, consider the limit $t \to -\infty$ and without loss of generality 
$g\equiv 1$. By the 
gauge 
transformations and boundary conditions one can find a (framed) gauge 
transformation such that
\begin{align}
 H_- = \Ad_{g_0}  (2Y_{n+1}) - \frac{\diff g_0}{\diff t} g_0^{-1}.
\end{align}
The complex equations for $Y''_j=Y_j^{g_0}$ then reduce to
\begin{align}
\begin{aligned}
 \frac{\diff}{\diff t} Y''_j + \frac{n+1}{n} Y''_j +  
\com{H^{-}}{Y''_j} &=0 \, ,\\
 \com{Y''_j}{Y''_k} &=0 \; ,
 \end{aligned}
\end{align}
for which the general solution is of the form
\begin{align}
 Y''_j(t)= e^{-\frac{n+1}{n}t} \Ad_{e^{(-H^{-} t)}}(\omega_j) \, ,
\end{align}
where the $\omega_j$ need to commute among each other. Note that the homomorphism $\rho_-$ induces a decomposition 
\begin{align}
\label{eq:decomp_rho}
 \glrmL(p,\C) = \bigoplus_{\vec{\mu} 
 \in \surmL(n), i \in \urmL(1) } V_{\vec{\mu},i} \, ,
\end{align}
where $\vec{\mu}$ labels representations of $\surm(n)$, while $i$ is the 
eigenvalue of $\ad_{H^-}$ corresponding to the label for the $\urm(1)$ 
centraliser of $\surm(n)$ inside $\surm(n{+}1)$.

Due to the boundary conditions we write $\omega_j = F_j^- + 
\delta_j$ and therefore obtain
\begin{align}
 Y''_j(t)= F_j^- + e^{-\frac{n+1}{n}t} \Ad_{e^{(-H^{-} t)}}(\delta_j) \; .
\end{align}
Additionally, we need to satisfy the commutator constraint in \eqref{eq:cplx_eq_CY} for which we find 
\begin{align}
 \com{F_j^-}{\delta_k} + \com{\delta_j}{F_k^-} =0 \; , \qquad
 \com{\delta_j}{\delta_k} =0 \; ,
\end{align}
using that $\com{F_{j}^{-}}{F_{k}^{-}}=0$ by the earlier arguments.
Moreover, the $\ad_{H^-}$-eigenvalues $i$ of $\delta_j$ are restricted by 
demanding that the $\delta_j$ contribution does not interfere with the 
boundary condition, i.e.\ 
\begin{align}
  \lim_{t\to -\infty} e^{-\frac{n+1}{n}t} \Ad_{e^{(-H^{-} t)}}(\delta_j) =0 
\qquad \Leftrightarrow \qquad
i(\delta_j) < - \frac{n+1}{n}.
\label{eq:cond_delta_Lem10}
\end{align}
The inequality here means that $\delta_j$ is an element of a subspace of the 
decomposition \eqref{eq:decomp_rho}, where the $\ad_{H^-}$-eigenvalue is 
bounded by the above expression.
For the case of the original Nahm equations, i.e.\ $n=1$, this yields the bound 
$i(\delta) < -2$ from \cite{Kronheimer:1990ay}.
There is still a remaining gauge freedom by 
\begin{align}
g= g_1 \cdot \ldots \cdot g_n \; , \qquad 
 g_j = e^{-H^- t} e^{\gamma_j} e^{H^- t} \; ,
\end{align}
which acts on all $Y'_j$ the same and where the contributions $g_j$ are suitably chosen such that they yield the desired action
on each element on the tuple. More precisely,
 the appearing $\gamma_j$ are restricted by 
demanding
\begin{enumerate}
 \item [(i)] $g$ preserves $2Y''_{n+1}=H^-$,
 \item [(ii)] $\lim_{t\to-\infty} g =1$,
 \item [(iii)] $g_j$ preserves $Y''_k = F_k^- + e^{-\frac{n+1}{n}t} \Ad_{e^{(-H^{-} 
t)}}(\delta_j) $ for all $k\neq j$.
\end{enumerate}
The first condition (i) is satisfied by noticing
\begin{align}
 g= e^{-H^- t} \left(  \prod_j e^{\gamma_j} \right) e^{H^- t} \; .
\end{align}
The third condition (iii) requires 
\begin{align}
 \gamma_j \in \bigcap_{k\neq j} z(F_k^-) \; , \qquad 
 \com{\gamma_j}{\delta_k} =0 = \com{\gamma_j}{\gamma_k} 
 \, , \quad j\neq k \, ,
\end{align}
while the second condition (ii) gives a restriction on the eigenvalues of 
$\gamma_j$
\begin{align}
 \lim_{t\to -\infty} g_j =1 \quad \Leftrightarrow \quad
 i(\gamma_j)<0 .
 \label{eq:cond_gamma_Lem10}
\end{align}
Applying the overall gauge transformations, we arrive at 
\begin{align}
 Y'''_k = F_k^- + e^{-\frac{n+1}{n}t} \Ad_{e^{(-H^{-} t)}} \left( 
\Ad_{e^{\gamma_j}} \left( F_j^-  +  \delta_j \right) - F_j^-  \right).
\end{align}
Similarly to \cite{Kronheimer:1990ay}, the only step left to prove is that for 
each $\delta_j$ there exists a $\gamma_j$ such that 
\begin{align}
 \Ad_{e^{\gamma_j}}\left( F_j^- + \delta_j  \right) - F_j^- =0 \; .
\end{align}
We have to prove a similar statement for the adaptation of Lemma 11 and provide 
the details there.

\paragraph{Adaptation of Lemma 11.}
Let $(Y_{n+1},Y_j)$ be a solution of the complex equations \eqref{eq:cplx_eq_CY}
satisfying the boundary conditions \eqref{eq:bc_+infty}. Then there exists a 
unique gauge transformation 
$g_+:\R \to \glrm(p,\C)$ 
with $\lim_{t\to \infty} g_+ = 1$ such that 
$(Y'_{n+1},Y'_j)=g_+(Y_{n+1},Y_j)$ satisfies
\begin{align}
 2 Y'_{n+1} = H^+= \mathrm{const} \;, \qquad Y'_j (t=0)\in S(F_j^+) \; .
 \end{align}
In other words,
\begin{align}
 (Y'_1 ,\ldots,Y'_n )(t=0) \in S_{\diag} \, ,
\end{align}
as defined in \eqref{eq:def_transvers_slice_diag}.
The first part of the generalisation of \cite[Lem.\ 11]{Kronheimer:1990ay} 
proceeds as above, the only changes are that the general solutions look
$Y''_k = F_k^+ + e^{-\frac{n+1}{n}t} \Ad_{e^{(-H^{+} 
t)}}(\epsilon_j) $ and the $\epsilon_j$ satisfy
\begin{align}
 \com{\epsilon_j}{\epsilon_k} =0 \;, \qquad 
 \com{F_j^+}{\epsilon_k} + \com{\epsilon_j}{F_k^+} =0  \;, \qquad 
 i(\epsilon_j) >-\frac{n+1}{n} \; .
 \label{eq:cond_epsilon_Lem11}
\end{align}
Note that the $i$ eigenvalues are with respect to $\ad_{H^+}$ and recall that we are now considering the opposite limit compared to Lemma 10
(and therefore have the opposite inequality).

In the next step, one applies the similar gauge transformations 
\begin{align}
g= g_1 \cdot \ldots \cdot g_n \; , \qquad 
 g_j = e^{-H^+ t} e^{\gamma_j} e^{H^+ t} \; ,
\end{align}
this time subject to the conditions
\begin{align}
 \gamma_j \in \bigcap_{k\neq j} z(F_k^+) \; , \qquad 
 \com{\gamma_j}{\epsilon_k} =0 = \com{\gamma_j}{\gamma_k} 
 \, , \quad j\neq k \, , \qquad
  i(\gamma_j)>0 \;.
  \label{eq:cond_gamma_Lem11}
\end{align}
Then one needs to prove that for each $\epsilon_j$ there exists a unique 
$\gamma_j$ such that 
\begin{align}
 \Ad_{e^{\gamma_j}}\left( F_j^+ + \epsilon_j  \right) - F_j^+ \in z(E_j^+) \; .
 \label{eq:proof_unique}
\end{align}
Again, the arguments by Kronheimer apply, but let us be more explicit.
The homomorphism $\rho_+$ induces the decomposition 
$  \glrmL(p,\C)|_{\slrmL(n{+}1,\C)} = \oplus_{\kappa} V_{\kappa}$
into $\slrmL(n{+}1,\C)$ irreps $V_\kappa$. These decompose further under 
$\Ad_{H^+}$ into $1$-dimensional irreps 
$ V_\kappa = \oplus_i V_{\kappa,i}$.
Since $V_\kappa$ is an $\slrmL(n{+}1,\C)$ irrep, the highest weight vector also 
has the maximal $i$ eigenvalue $\lambda_\kappa>0$ of all weight 
vectors of $V_\kappa$.

The linearisation of \eqref{eq:proof_unique} is $\epsilon_j - 
\com{F_j^+}{\gamma_j} \in z(E_j^+)$. Since $F_j^+$ is a annihilation operator, 
it decreases all $i$ eigenvalues by a certain increment $\Delta_j$; in other 
words $\ad_{F_j^+} : \oplus_{i>0} V_i \to \oplus_{i>-\Delta_j} V_i$, where 
$V_i$ are eigenspaces with certain $i$ eigenvalue. This map is injective, 
because the kernel of $\ad_{F_j^+} $ is not contained in the domain since all 
$i>0$. Moreover, the image lies in the complement of $z(E_j^+)$, because all $i$ 
eigenvalues have been lowered by $\Delta_j$, therefore none of the weight 
vectors 
has the maximal eigenvalue and cannot be annihilated by $E_j^+$. Consequently, 
for any $\epsilon_j$ one can find a unique $\gamma_j$ to match the part of 
$\epsilon_j$ in the orthogonal complement of $z(E_j^+)$; hence, the claim holds.

\paragraph{Both Lemmata together.}
Consequently, a complex trajectory is equivalent to a tuple 
$(Y_{n+1},Y_j)$ satisfying the conditions
\begin{alignat}{3}
 Y_{n+1}(t) &= \frac{1}{2} H^- \;, \qquad &
 Y_{j}(t) &= F_j^- \; , \qquad & 
 t &\in (-\infty,0] \; , \\
 Y_{n+1}(t) &= \frac{1}{2} H^+ \;, \qquad &
 Y_{j}(t) &= F_j^+ + e^{-\frac{n+1}{n}t} 
\Ad_{e^{-H^+ t}} (\epsilon_j) \; , \qquad  &
t &\in [1,\infty) \; . 
\end{alignat}
The choice of $\epsilon_j$ is such that $(F_1^+ + \epsilon_1, \ldots, F_n^+ + 
\epsilon_n) \in S_{\diag}(\rho_+)$. Since the solution is locally constant, it 
follows that there exists an $g\in \glrm(p,\C)$ such that
\begin{align}
 \Ad_g (F_1^- , \ldots, F_n^-)= (F_1^+ + \epsilon_1, \ldots , F_n^+ + 
\epsilon_n) \in \Ncal_{\diag}(\rho_-) \; .
\end{align}
Hence, the complex trajectories are classified by the intersection 
$\Ncal_{\diag}(\rho_-) \cap S_{\diag}(\rho_+)$.
%

\bibliographystyle{JHEP}
\bibliography{Bibliography6}

\providecommand{\href}[2]{#2}\begingroup\raggedright\begin{thebibliography}{10}

\bibitem{Gross:2003}
M.~Gross, D.~Huybrechts, and D.~Joyce, {\em {Calabi-{Y}au manifolds and related
  geometries}}.
\newblock Universitext. Springer-Verlag, Berlin, 2003.
\newblock Lectures from the Summer School held in Nordfjordeid, June 2001.

\bibitem{Hitchin:1986ea}
N.~J. Hitchin, A.~Karlhede, U.~Lindstrom, and M.~Rocek, {\it {Hyperk\"ahler
  Metrics and Supersymmetry}},  {\em Commun. Math. Phys.} {\bf 108} (1987) 535.

\bibitem{Gibbons:1998xa}
G.~W. Gibbons and P.~Rychenkova, {\it {Cones, tri-Sasakian structures and
  superconformal invariance}},  {\em Phys. Lett.} {\bf B443} (1998) 138--142,
  [\href{http://arxiv.org/abs/hep-th/9809158}{{\tt hep-th/9809158}}].

\bibitem{deWit:1998zg}
B.~de~Wit, B.~Kleijn, and S.~Vandoren, {\it {Rigid N=2 superconformal
  hypermultiplets}},  \href{http://arxiv.org/abs/hep-th/9808160}{{\tt
  hep-th/9808160}}. [Lect. Notes Phys.524,37(1999)].

\bibitem{Donaldson:1983}
S.~K. Donaldson, {\it {Self-dual connections and the topology of smooth
  {$4$}-manifolds}},  {\em Bull. Amer. Math. Soc. (N.S.)} {\bf 8} (1983)
  81--83.

\bibitem{Corrigan:1982th}
E.~Corrigan, C.~Devchand, D.~B. Fairlie, and J.~Nuyts, {\it {First Order
  Equations for Gauge Fields in Spaces of Dimension Greater Than Four}},  {\em
  Nucl. Phys.} {\bf B214} (1983) 452--464.

\bibitem{Donaldson:1985}
S.~K. Donaldson, {\it {Anti self-dual Yang-Mills connections over complex
  algebraic surfaces and stable vector bundles}},  {\em Proceedings of the
  London Mathematical Society} {\bf 3} (1985) 1--26.

\bibitem{Uhlenbeck:1986}
K.~Uhlenbeck and S.-T. Yau, {\it {On the existence of Hermitian-Yang-Mills
  connections in stable vector bundles}},  {\em Communications on Pure and
  Applied Mathematics} {\bf 39} (1986).

\bibitem{Salamon:1988}
M.~Mamone~Capria and S.~M. Salamon, {\it {Yang-Mills fields on quaternionic
  spaces}},  {\em Nonlinearity} {\bf 1} (1988) 517--530.

\bibitem{Nitta:1988}
T.~Nitta, {\it {Vector bundles over quaternionic {K}\"ahler manifolds}},  {\em
  Tohoku Math. J. (2)} {\bf 40} (1988) 425--440.

\bibitem{Bartocci:2004}
C.~Bartocci and M.~Jardim, {\it {Hyperk{\"a}hler Nahm transforms}},  in {\em
  CRM Proc. Lecture Notes}, vol.~38, pp.~103--111, 2004.

\bibitem{Harland:2009yu}
D.~Harland, T.~A. Ivanova, O.~Lechtenfeld, and A.~D. Popov, {\it {Yang-Mills
  flows on nearly K\"ahler manifolds and G(2)-instantons}},  {\em Commun. Math.
  Phys.} {\bf 300} (2010) 185--204, [\href{http://arxiv.org/abs/0909.2730}{{\tt
  arXiv:0909.2730}}].

\bibitem{Harland:2010ix}
D.~Harland and A.~D. Popov, {\it {Yang-Mills fields in flux compactifications
  on homogeneous manifolds with SU(4)-structure}},  {\em JHEP} {\bf 02} (2012)
  107, [\href{http://arxiv.org/abs/1005.2837}{{\tt arXiv:1005.2837}}].

\bibitem{Bauer:2010fia}
I.~Bauer, T.~A. Ivanova, O.~Lechtenfeld, and F.~Lubbe, {\it {Yang-Mills
  instantons and dyons on homogeneous $G_2$-manifolds}},  {\em JHEP} {\bf 10}
  (2010) 044, [\href{http://arxiv.org/abs/1006.2388}{{\tt arXiv:1006.2388}}].

\bibitem{Haupt:2011mg}
A.~S. Haupt, T.~A. Ivanova, O.~Lechtenfeld, and A.~D. Popov, {\it {Chern-Simons
  flows on Aloff-Wallach spaces and Spin(7)-instantons}},  {\em Phys. Rev.}
  {\bf D83} (2011) 105028, [\href{http://arxiv.org/abs/1104.5231}{{\tt
  arXiv:1104.5231}}].

\bibitem{Gemmer:2011cp}
K.-P. Gemmer, O.~Lechtenfeld, C.~N{\"o}lle, and A.~D. Popov, {\it {Yang-Mills
  instantons on cones and sine-cones over nearly Kahler manifolds}},  {\em
  JHEP} {\bf 09} (2011) 103, [\href{http://arxiv.org/abs/1108.3951}{{\tt
  arXiv:1108.3951}}].

\bibitem{harland_noelle}
D.~Harland and C.~N{\"o}lle, {\it {Instantons and Killing spinors}},  {\em
  JHEP} {\bf 03} (2012) 082, [\href{http://arxiv.org/abs/1109.3552}{{\tt
  arXiv:1109.3552}}].

\bibitem{ivanova_popov}
T.~A. Ivanova and A.~D. Popov, {\it {Instantons on special holonomy
  manifolds}},  {\em Phys. Rev.} {\bf D85} (2012) 105012,
  [\href{http://arxiv.org/abs/1203.2657}{{\tt arXiv:1203.2657}}].

\bibitem{Bunk:2014kva}
S.~Bunk, T.~A. Ivanova, O.~Lechtenfeld, A.~D. Popov, and M.~Sperling, {\it
  {Instantons on sine-cones over Sasakian manifolds}},  {\em Phys. Rev.} {\bf
  D90} (2014) 065028, [\href{http://arxiv.org/abs/1407.2948}{{\tt
  arXiv:1407.2948}}].

\bibitem{Bunk:2014coa}
S.~Bunk, O.~Lechtenfeld, A.~D. Popov, and M.~Sperling, {\it {Instantons on
  conical half-flat 6-manifolds}},  {\em JHEP} {\bf 01} (2015) 030,
  [\href{http://arxiv.org/abs/1409.0030}{{\tt arXiv:1409.0030}}].

\bibitem{Sperling:2015sra}
M.~Sperling, {\it {Instantons on Calabi-Yau cones}},  {\em Nucl. Phys.} {\bf
  B901} (2015) 354--381, [\href{http://arxiv.org/abs/1505.01755}{{\tt
  arXiv:1505.01755}}].

\bibitem{Haupt:2015wdq}
A.~S. Haupt, {\it {Yang-Mills solutions and Spin(7)-instantons on cylinders
  over coset spaces with $G_2$-structure}},  {\em JHEP} {\bf 03} (2016) 038,
  [\href{http://arxiv.org/abs/1512.07254}{{\tt arXiv:1512.07254}}].

\bibitem{Gaiotto:2008sa}
D.~Gaiotto and E.~Witten, {\it {Supersymmetric Boundary Conditions in N=4 Super
  Yang-Mills Theory}},  {\em J. Statist. Phys.} {\bf 135} (2009) 789--855,
  [\href{http://arxiv.org/abs/0804.2902}{{\tt arXiv:0804.2902}}].

\bibitem{Gaiotto:2008ak}
D.~Gaiotto and E.~Witten, {\it {S-Duality of Boundary Conditions In N=4 Super
  Yang-Mills Theory}},  {\em Adv. Theor. Math. Phys.} {\bf 13} (2009), no.~3
  721--896, [\href{http://arxiv.org/abs/0807.3720}{{\tt arXiv:0807.3720}}].

\bibitem{Xie:2013gma}
D.~Xie, {\it {M5 brane and four dimensional N = 1 theories I}},  {\em JHEP}
  {\bf 04} (2014) 154, [\href{http://arxiv.org/abs/1307.5877}{{\tt
  arXiv:1307.5877}}].

\bibitem{Heckman:2016xdl}
J.~J. Heckman, P.~Jefferson, T.~Rudelius, and C.~Vafa, {\it {Punctures for
  theories of class $ {\mathcal{S}}_{\varGamma } $}},  {\em JHEP} {\bf 03}
  (2017) 171, [\href{http://arxiv.org/abs/1609.01281}{{\tt arXiv:1609.01281}}].

\bibitem{Hashimoto:2014vpa}
A.~Hashimoto, P.~Ouyang, and M.~Yamazaki, {\it {Boundaries and defects of $
  \mathcal{N}=4 $ SYM with 4 supercharges. Part I: Boundary/junction
  conditions}},  {\em JHEP} {\bf 10} (2014) 107,
  [\href{http://arxiv.org/abs/1404.5527}{{\tt arXiv:1404.5527}}].

\bibitem{Hashimoto:2014nwa}
A.~Hashimoto, P.~Ouyang, and M.~Yamazaki, {\it {Boundaries and defects of $
  \mathcal{N}=4 $ SYM with 4 supercharges. Part II: Brane constructions and 3d
  $ \mathcal{N}=2 $ field theories}},  {\em JHEP} {\bf 10} (2014) 108,
  [\href{http://arxiv.org/abs/1406.5501}{{\tt arXiv:1406.5501}}].

\bibitem{boyer_galicki}
C.~P. Boyer and K.~Galicki, {\it {3 - Sasakian manifolds}},  {\em Surveys Diff.
  Geom.} {\bf 7} (1999) 123--184,
  [\href{http://arxiv.org/abs/hep-th/9810250}{{\tt hep-th/9810250}}].

\bibitem{boyer_galicki_mann}
C.~P. Boyer, K.~Galicki, and B.~M. Mann, {\it {The geometry and topology of
  3-Sasakian manifolds}},  {\em J. reine angew. Math} {\bf 455} (1994)
  183--220.

\bibitem{hitchin_hyperkaehler}
N.~Hitchin, {\it {Hyperk\"ahler manifolds}},  {\em Séminaire Bourbaki} {\bf
  34} (1991-1992) 137--166.

\bibitem{BFGK}
H.~Baum, T.~Friedrich, R.~Grunewald, and I.~Kath, {\em {Twistors and Killing
  spinors on Riemannian manifolds}}.
\newblock Teubner-Texte zur Mathematik, 1991.

\bibitem{Atiyah:1982fa}
M.~F. Atiyah and R.~Bott, {\it {The Yang-Mills equations over Riemann
  surfaces}},  {\em Phil. Trans. Roy. Soc. Lond.} {\bf A308} (1982) 523--615.

\bibitem{Deser:2014zya}
A.~Deser, O.~Lechtenfeld, and A.~D. Popov, {\it {Sigma-model limit of
  Yang–Mills instantons in higher dimensions}},  {\em Nucl. Phys.} {\bf B894}
  (2015) 361--373, [\href{http://arxiv.org/abs/1412.4258}{{\tt
  arXiv:1412.4258}}].

\bibitem{s7}
J.~C. Geipel, O.~Lechtenfeld, A.~D. Popov, and R.~J. Szabo, {\it {Sasakian
  quiver gauge theories and instantons on cones over round and squashed
  seven-spheres}},  \href{http://arxiv.org/abs/1706.07383}{{\tt
  arXiv:1706.07383}}.

\bibitem{Lechtenfeld:2008nh}
O.~Lechtenfeld, A.~D. Popov, and R.~J. Szabo, {\it {SU(3)-equivariant quiver
  gauge theories and nonabelian vortices}},  {\em JHEP} {\bf 08} (2008) 093,
  [\href{http://arxiv.org/abs/0806.2791}{{\tt arXiv:0806.2791}}].

\bibitem{Dolan:2010ur}
B.~P. Dolan and R.~J. Szabo, {\it {Equivariant dimensional reduction and quiver
  gauge theories}},  {\em Gen. Rel. Grav.} {\bf 43} (2010) 2453,
  [\href{http://arxiv.org/abs/1001.2429}{{\tt arXiv:1001.2429}}].

\bibitem{Lechtenfeld:2015ona}
O.~Lechtenfeld, A.~D. Popov, M.~Sperling, and R.~J. Szabo, {\it {Sasakian
  quiver gauge theories and instantons on cones over lens 5-spaces}},  {\em
  Nucl. Phys.} {\bf B899} (2015) 848--903,
  [\href{http://arxiv.org/abs/1506.02786}{{\tt arXiv:1506.02786}}].

\bibitem{Geipel:2016uij}
J.~C. Geipel, O.~Lechtenfeld, A.~D. Popov, and R.~J. Szabo, {\it {Sasakian
  quiver gauge theories and instantons on the conifold}},  {\em Nucl. Phys.}
  {\bf B907} (2016) 445--475, [\href{http://arxiv.org/abs/1601.05719}{{\tt
  arXiv:1601.05719}}].

\bibitem{Geipel:2016hpk}
J.~C. Geipel, {\it {Sasakian quiver gauge theory on the Aloff–Wallach space
  $X_{1,1}$}},  {\em Nucl. Phys.} {\bf B916} (2017) 279--303,
  [\href{http://arxiv.org/abs/1605.03521}{{\tt arXiv:1605.03521}}].

\bibitem{kronheimer1990hyper}
P.~B. Kronheimer, {\it {A hyper-K\"ahlerian structure on coadjoint orbits of a
  semisimple complex group}},  {\em Journal of the London Mathematical Society}
  {\bf 2} (1990) 193--208.

\bibitem{Kronheimer:1990ay}
P.~B. Kronheimer, {\it {Instantons and the geometry of the nilpotent variety}},
   {\em J. Diff. Geom.} {\bf 32} (1990) 473--490.

\bibitem{donaldson1984}
S.~K. Donaldson, {\it Nahm's equations and the classification of monopoles},
  {\em Comm. Math. Phys.} {\bf 96} (1984) 387--407.

\bibitem{Lechtenfeld:2012yw}
O.~Lechtenfeld, {\it {Instantons and Chern-Simons flows in 6, 7 and 8
  dimensions}},  {\em Phys. Part. Nucl.} {\bf 43} (2012) 569--576,
  [\href{http://arxiv.org/abs/1201.6390}{{\tt arXiv:1201.6390}}].

\bibitem{Sperling:2016dqk}
M.~Sperling, {\em {Two aspects of gauge theories : higher-dimensional
  instantons on cones over Sasaki-Einstein spaces and Coulomb branches for
  3-dimensional N=4 gauge theories}}.
\newblock PhD thesis, Hannover U., 2016.

\bibitem{Ginzburg:2000}
V.~Ginzburg, {\it {Principal nilpotent pairs in a semisimple {L}ie algebra.
  {I}}},  {\em Invent. Math.} {\bf 140} (2000), no.~3 511--561,
  [\href{http://arxiv.org/abs/math/9903059}{{\tt math/9903059}}].

\bibitem{Panyushev:2000}
D.~I. Panyushev, {\it {Nilpotent pairs in semisimple {L}ie algebras and their
  characteristics}},  {\em Internat. Math. Res. Notices} (2000), no.~1 1--21,
  [\href{http://arxiv.org/abs/math/9906049}{{\tt math/9906049}}].

\bibitem{Panyushev:2001}
D.~I. Panyushev, {\it {Nilpotent pairs, dual pairs, and sheets}},  {\em J.
  Algebra} {\bf 240} (2001), no.~2 635--664,
  [\href{http://arxiv.org/abs/math/9904014}{{\tt math/9904014}}].

\bibitem{Elashvili:2001}
A.~G. Elashvili and D.~I. Panyushev, {\it {A classification of the principal
  nilpotent pairs in simple {L}ie algebras and related problems}},  {\em J.
  London Math. Soc. (2)} {\bf 63} (2001), no.~2 299--318,
  [\href{http://arxiv.org/abs/math/9909082}{{\tt math/9909082}}].

\bibitem{Biquard:1996}
O.~Biquard, {\it {Sur les \'equations de {N}ahm et la structure de {P}oisson
  des alg\`ebres de {L}ie semi-simples complexes}},  {\em Math. Ann.} {\bf 304}
  (1996) 253--276.

\bibitem{kovalev}
A.~G. Kovalev, {\it {Nahm's equations and complex adjoint orbits}},  {\em
  Quart. J. Math. Oxford Ser. (2)} {\bf 47} (1996), no.~185 41--58.

\bibitem{Fuchs:1997jv}
J.~Fuchs and C.~Schweigert, {\em {Symmetries, Lie algebras and representations:
  A graduate course for physicists}}.
\newblock Cambridge University Press, 2003.

\end{thebibliography}\endgroup
\end{document}